\documentclass[english]{article}
\usepackage{helvet}
\usepackage[T1]{fontenc}
\usepackage{a4wide}
\usepackage{babel}
\usepackage{graphics}
\usepackage{subfigure}

\makeatletter

\providecommand{\LyX}{L\kern-.1667em\lower.25em\hbox{Y}\kern-.125emX\@}
\newcommand{\noun}[1]{\textsc{#1}}
\let\SF@@footnote\footnote
\def\footnote{\ifx\protect\@typeset@protect
    \expandafter\SF@@footnote
  \else
    \expandafter\SF@gobble@opt
  \fi
}
\expandafter\def\csname SF@gobble@opt \endcsname{\@ifnextchar[
  \SF@gobble@twobracket
  \@gobble
}
\edef\SF@gobble@opt{\noexpand\protect
  \expandafter\noexpand\csname SF@gobble@opt \endcsname}
\def\SF@gobble@twobracket[#1]#2{}

 \usepackage{verbatim}

\usepackage{lyx}
\usepackage{rawfonts}
\newcommand{\dittocaption}{%
   \renewcommand{\addcontentsline}[3]{}
   \addtocounter{figure}{-1}
   \caption
}

\newcommand{\ltae}{\raisebox{-0.6ex}{$\,\stackrel
{\raisebox{-.2ex}{$\textstyle <$}}{\sim}\,$}}
\newcommand{\gtae}{\raisebox{-0.6ex}{$\,\stackrel
{\raisebox{-.2ex}{$\textstyle >$}}{\sim}\,$}}

  \DeclareSymbolFont{UPM}{U}{eur}{m}{n}
  \SetSymbolFont{UPM}{bold}{U}{eur}{b}{n}
  \DeclareMathSymbol{\upi}{0}{UPM}{"19}
  \DeclareMathSymbol{\umu}{0}{UPM}{"16}
  \DeclareMathSymbol{\upartial}{0}{UPM}{"40}

\makeatother

\begin{document}

\title{Testing the photoionization models of powerful radio galaxies: Mixed line-emitting
media in 3C 321}

\author{T. G. Robinson\protect\( ^{1,2}\protect \), C. N. Tadhunter\protect\( ^{2}\protect \),
D. J. Axon\protect\( ^{3}\protect \) \& A. Robinson\protect\( ^{3}\protect \)}

\maketitle
\( ^{1} \)E-mail: t.g.robinson@shef.ac.uk

\( ^{2} \)Department of Physics and Astronomy, University of Sheffield, Sheffield,
S7 3RH.

\( ^{3} \)Department of Physical Sciences, University of Hertfordshire, Hatfield,
Hertfordshire, AL10 9AB

\begin{abstract}
The photoionization models for the narrow emission line regions of powerful
radio galaxies have yet to be tested in depth. To this end, we present high-quality
long-slit spectroscopy of the powerful double-nucleus radio galaxy 3C 321. The
data have good enough spatial resolution to be able to trace the variation in
emission-line properties on kpc scales. Continuum modelling and subtraction
enables the faint emission line fluxes to be measured in several regions across
the emission line nebula. We plot diagnostic line-ratio diagrams and compare
them with the predictions of various photoionization models, finding that the
data is best fit by models which assume a mixture of optically thin and thick
clouds illuminated by a power-law continuum. 

The emission line kinematics, line ratios and deduced physical conditions show
remarkably little variation across the source. We deduce a mean electron density
of \( 400\pm 120 \) \( \mbox {cm}^{-3} \) and a mean temperature of \( 11500\pm 1500 \)
K. Assuming a single population of optically thick line-emitting clouds, we
calculate a mean photoionization parameter of \( (1.1\pm 0.5)\times 10^{-2} \)
and hence a photoionizing photon luminosity of \( Q\sim 10^{55}-10^{56}\mbox {phots}^{-1}\mbox {sr}^{-1} \).
This indicates a central engine as luminous as that of the powerful quasar 3C
273, yet there is no evidence for such an energetically prolific central engine
at either far-infrared or radio wavelengths. We therefore conclude that the
mixed-media models, which give \( Q\sim 5\times 10^{53}-5\times 10^{54} \),
represent a more likely scenario. 

As a by-product of the continuum subtraction we infer that young stellar populations
account for \( \sim 0.4\% \) of the visible stellar mass in the galaxy, and
that these populations are spatially extended.
\end{abstract}
\textbf{Key words:} galaxies: active -- galaxies: individual: 3C 321 -- galaxies:
stellar content -- galaxies: quasars: emission lines -- line: formation

\section{Introduction}

Emission lines are the most easily measured features of radio galaxies, and
yet the physical mechanisms involved in their production are not well understood.
\textbf{\noun{}}Various suggestions have been made as to the method of ionization
of the line-emitting gas, including shocks from interactions between the radio
jets and the warm clouds \cite{Clark+A+T-1998} and photoionization by a central
source \cite{Robinson+V-V+A+etal-1994}. Central source photoionization is particularly
attractive in light of unified schemes for powerful radio sources, which hold
that radio galaxies and quasars are a single class of object viewed from different
directions \textbf{\noun{}}\cite{Barthel-1989,Lawrence-1991}. If these orientation-dependent
unified schemes are true, and radio galaxies contain quasar nuclei obscured
from direct observation by a dusty torus, we expect to see the effects of the
powerful AGN illuminating the surrounding regions. Assuming that central source
photoionization is the dominant mechanism, the emission line spectra may be
used to infer the properties of the central engine, such as the shape of the
photoionizing continuum and its power \cite{Tadhunter+M+R-1994,Morganti+R+F+etal-1991}.
Indeed, emission-line luminosity has been shown to correlate broadly with 178
MHz radio power \cite{Tadhunter+M+R+D+V-M+F-1998,Rawlings+S-1991}, suggesting
a link between the AGN continuum emission and the mechanism of radio-jet production.

One way of investigating the properties of the central engine is by modelling
the emission line spectrum of the photoionized gas \cite{Robinson+B+F+T-1987,Binette+R+C-1988}.
In general, these models depend on both the properties of the line-emitting
medium and the ionizing spectrum, which in turn depends on the properties of
the central AGN. The traditional approach has been to assume a power-law or
hot black-body continuum ionizing optically thick, solar abundance clouds of
gas. This approach has been successful up to a point, but there still remain
several outstanding difficulties in explaining the observed spectra of radio
galaxies in terms of pure photoionization:

\begin{itemize}
\item \textbf{Discrepant line ratios.} Some line ratios are problematic in that they
are not correctly predicted by photoionization models in some objects. These
include {[}OIII{]}\( \lambda 5007 \)/\( \lambda 4363 \) \cite{Tadhunter+F+Q-1989,Morganti+T+D+S-1997},
HeII\( \lambda 4686 \)/H\( \beta  \) \cite{Binette+W+S-B-1996} and {[}NII{]}\( \lambda 6583 \)/{[}OII{]}\( \lambda 3727 \)
\cite{Tadhunter+M+R-1994}. At least some of these discrepancies can be explained
in terms of departures from the assumptions of the simplest models. For example,
the discrepant {[}OIII{]}\( \lambda 5007 \)/\( \lambda 4363 \), HeII\( \lambda 4686 \)/H\( \beta  \)
ratios can be modelled if the photoionized medium is a mixture of optically
thin and thick components \cite{Binette+W+S-B-1996}, while the {[}NII{]}\( \lambda 6583 \)/{[}OII{]}\( \lambda 3727 \)
can be explained in terms of non-solar abundances \cite{Viegas+P-1992}. However,
the mixed-medium model requires specific column depths in the optically thin
and thick clouds and cannot explain some of the more extreme line ratios measured
in some sources.
\item \textbf{Spatial variations in line ratios.} These are often difficult to explain
in terms of a central ionizing source. For example, the {[}NII{]}\( \lambda 6584 \)/H\( \alpha  \)
and {[}SII{]}\( \lambda 6726 \)/H\( \alpha  \) line ratios in Cygnus A show
no variation in value across both core and extended regions, despite the fact
that {[}OIII{]} line ratios vary dramatically across the same regions \cite{Tadhunter+M+R-1994}.
\item \textbf{The effects of shocks.} Line broadening and differences in the widths
of lines of different ionization suggest shocks are also important, even in
the narrow line region (NLR) where there is little direct morphological evidence
for jet/cloud interaction \cite{Clark+A+T-1998,Villar-Martin+T+M+etal-1999,Capetti+A+M+etal-1999}.
\end{itemize}
To test the photoionization models, high quality spatially resolved spectra
of a predominantly photoionized radio galaxy are required. We have obtained
such data for the powerful radio galaxy 3C 321. The advantages of this dataset
are that it has wide spectral coverage, accurate flux calibration, low airmass,
subarcsecond seeing, S/N ratios high enough to measure electron density in several
spatial bins (a feature not shared by any other radio galaxy with published
spectra), and a high enough S/N in the continuum to allow accurate modelling
and subtraction.

The double nucleus host galaxy of 3C 321 is an ideal object for an in-depth
study of photoionization models. Its relatively low redshift (\( z=0.0961 \))
means that ground-based long-slit spectroscopy allows the study of the variation
of its emission-line spectrum on kpc scales. 20cm VLA radio images in combination
with emission line images \cite{Baum+H+B+vB+M-1988} show little sign of jet-cloud
interaction, despite the overall complexity of the emission line structure,
and there is little evidence for the extreme emission-line kinematics which
would indicate the action of shocks, as observed, for example, in 3C 171 \cite{Clark+A+T-1998}.
Polarization studies also provide evidence of illumination by a central source,
since scattered quasar features are detected in both polarized and direct light
\cite{Filippenko-1987,Draper+S+T-1993,Young+H+E+etal-1996}. Furthermore, the
fact that 3C 321 is at the brighter end of the 60 \( \mu  \)m luminosity range
of Narrow Line Radio Galaxies (NLRGs) \cite{Hes+B+H-1995,Heisler+V-1994} also
suggests that its central regions contain a powerful illuminating source\textbf{\emph{.}}
In addition to the quasar component, continuum modelling by \shortcite{Tadhunter+D+S-1996}
(TDS96) showed evidence for a significant starburst component. 

We present the long-slit data for 3C 321 in Sections \ref{sec: Observations}
\& \ref{sec: Results} and discuss the implications of these data for the photoionization
models of powerful radio galaxies in Section \ref{sec: Discussion}.

\section{Observations}

\label{sec: Observations}

\subsection{Data acquisition and reduction }

\label{sec: Data reduction} The observations were made in July 1996 using the
ISIS spectrograph on the 4.2m William Herschel Telescope on La Palma. Two sets
of 1500s exposures were taken simultaneously on the red and blue arms of ISIS
giving a wavelength coverage of 3630--9820 \AA. Not all of this range was useful,
as the dichroic in ISIS attenuates the region above \textbf{\emph{\( \sim 5900 \)}}
\AA~in the blue spectrum, and flux calibration errors dominate above \( \sim 8200 \)
\AA. Gaussian fits to the night sky lines show that the mean spectral resolution
of our data is \( 8.2\pm 0.1 \) \AA~in the blue spectra and \( 7.9\pm 0.9 \)
\AA~in the red.

The seeing was measured to be 0.8 arcsec (FWHM), and the low airmass (1.008)
ensured that the effects of differential atmospheric refraction were negligible
--- an important consideration given the large wavelength range covered by the
data. The spatial scale of our spectra is \( 0.79 \) kpc per pixel\footnote{%
Values of \( q_{0}=0.5 \) and \( \mbox {H}_{0}=50 \) km s\( ^{-1} \)Mpc\( ^{-1} \)
are assumed throughout.
}.

The slit was placed at a position angle of 130\( ^{\circ } \) on the sky, the
direction of the axis defined by the two nuclei of the object. Figure \ref{fig: HST Image}
shows a diagram of the slit superimposed on an {[}OIII{]}\( \lambda \lambda 4959+5007 \)~
HST image taken with WFPC2 using a linear ramp filter \cite{Martel+B+S+etal-1999}.
A clear dust lane is visible in the SE blob, which is not resolved in our long
slit images, although there is evidence for a significant increase in the E(B-V)
reddening in this region (Region F in Figure \ref{fig: HST Image} - see Table
\ref{tab: EL fluxes}). Also marked are the 7 regions chosen for the detailed
analysis described below. 

The data were reduced using the Starlink \noun{figaro} package and IRAF. Bias
subtraction was carried out and cosmic rays removed. The spectra were then wavelength-calibrated,
corrected for atmospheric extinction and flux calibrated, using the standard
stars and arc exposures taken at the same time as the science images. 
The data were also corrected for both S-distortion and tilt, which may be caused
by the slight misalignment of the slit with the grating and CCD, by using the
spectra of standard stars taken at the same time as the observations. After
correction, the residual offset in the spatial direction between the two ends
of the spectra was estimated to be \( <0.3 \) pixels. The two blue and two
red spectra were co-added and the composite blue spectrum was then re-sampled
to give the same spatial pixel scale as the red, and the region of the spectrum
severely attenuated by the dichroic was removed. Finally, the two composite
spectra were aligned and shifted back to the rest-frame using the mean redshift
measured from the brighter emission lines.

The reduction process resulted in two spectra with useful rest wavelengths in
the ranges \( 3314 \)--\( 5400 \) \& \( 5891 \)--\( 8200 \) \AA. We estimate
a flux calibration error of \( 5\% \) across this range.

\section{Results}

\label{sec: Results} Some analysis was made of the data prior to the continuum
subtraction. In particular, kinematics of the {[}OIII{]}\( \lambda 5007 \)
line were measured across the object on a pixel-by-pixel basis. The results
of this line-fitting are presented in section \ref{sec: Kinematics}.

Next, seven spatial bins were chosen for detailed analysis (see Figure \ref{fig: HST Image}).
These were selected to have some morphological meaning and/or to give a high
enough signal-to-noise in the density diagnostic {[}SII{]}\( \lambda \lambda 6717,6731 \)
lines to measure densities \( 3\sigma  \) away from the low-density limit.
The HST image was used to identify spatial bins which fall either side of the
central dust-lane, and to sub-divide the emission line and continuum spatial
peaks whilst maintaining a high signal-to-noise (see Figure \ref{fig: Bins}).
\textbf{\emph{}}Note that Gaussian fits to the mean spatial profiles of the
{[}OIII{]}\( \lambda 5007 \) and line-free continuum (\( \lambda 8849 \))
show significant offsets between the continuum and emission line peaks in both
nuclei, with the north-west emission line peak offset by \( 0.11\pm 0.04 \)''
(\( 0.26 \) kpc) to the north-west of the north-west continuum peak and the
south-east emission line peak offset by \( 0.66\pm 0.02 \)'' (\( 1.6 \) kpc)
to the SE of the SE continuum peak (Figure \ref{fig: Bins}).

\subsection{Continuum modelling}

\label{sec: Continuum subtraction} To enable the fainter emission lines to
be measured more accurately, model continua were calculated and subtracted from
the observed spectra. A nebular continuum was generated for each bin, comprising
the blended higher (>H8) Balmer series together with a theoretical nebular recombination
continuum (generated using \noun{dipso}), both of which were normalized to
the flux in H\( \beta  \). This was then convolved with the H\( \beta  \)
line width and subtracted from the data prior to the modelling (see \citeN{Dickson+T+S+etal-1995}).

Initially, the continuum in regions A-G was modelled using 2 components chosen
to represent AGN light (either direct or scattered) and the light of the host
elliptical galaxy: a power law of the form \( f_{\lambda }\propto \lambda ^{+\alpha } \)
(corresponding to a frequency power-law, \( \nu ^{-(\alpha +2)} \)) and a 15
Gyr old elliptical galaxy template taken from \shortcite{Bruzual+C-1993}. A
more sophisticated model was then introduced, following TDS96, this time with
an extra component corresponding to a contribution from a 0.1 Gyr starburst.
The flux in 36 continuum bins, selected from both the red and blue spectra,
was measured for the data and the model components, avoiding emission lines,
atmospheric absorption bands and image defects. The models were then generated
by choosing a normalizing continuum bin and scaling the different model components
so that the total model flux in the normalizing bin was less than 125\% of the
observed flux. In this way, a series of models was created, filling the 3- and
4-d parameter space defined by the relative fractions of each of the components,
and the value of \( \alpha  \). The best-fitting model was determined using
a reduced \( \chi  \)-squared test, with the fitting being performed by IDL
routines written for this purpose. The contributions of the different components
were varied between 0 -- 125\% of the flux in the normalizing bin in increments
of 2.5\%, and the power law was varied over the range \( -4.9<\alpha <5.1 \)
in 50 increments. The results of the modelling are shown in Tables \ref{tab: 2-component}
\& \ref{tab: Continuum components}. These were used to generate and then subtract
the best-fit model continuum from the observational data. The best fitting models
for the 3-component fits are plotted in Figure \ref{fig: Continuum subtracted spectra},
together with the nebular-component-subtracted continua of the different spatial
regions. \noun{}

We exclude region A from the discussion which follows. As Tables \ref{tab: 2-component}
\& \ref{tab: Continuum components} show, the value of \( \chi ^{2} \) is much
higher for this region than for the other regions. This is due the fact that
the continuum in region A is both weak and noisy, and the background subtraction
more uncertain. We therefore refrain from drawing any conclusions regarding
the contributions of individual continuum components in this region, although
we do subtract the best-fit continuum from the spectrum for the purposes of
studying the emission-line spectrum. 

The 2-component fits are entirely adequate for the regions containing the two
continuum peaks (C and F). For all other regions, however, the 2-component fits
are poor, with the data exhibiting an excess of \( 10-15 \) percent of the
continuum flux relative to the best-fit model in the wavelength region \( 3750 \)
-- \( 4300 \) \AA. A similar discrepancy was noted by TDS96 for the integrated
light of the nuclear regions. Moreover, the 2-component models also give an
over-prediction of the flux above \( \sim 7000 \) \AA. Note that the increased
red-coverage in our spectra compared with those of TDS96 gives us a `longer
lever' in terms of constraining the fits to the data.

The 3-component models fit the data substantially better, giving much lower
values of \( \chi _{min,reduced}^{2} \) and also giving a more convincing overall
spectral shape. As suggested by the 2-component models there is no strong evidence
for a young stellar population in regions C \& F, but the other four regions
have \( \gtae 20\% \) of their continuum at \( 5060 \) \AA ~contributed by
a starburst component. Note that the negative slope of the continuum between
\( 6500 \) \AA ~and \( 9000 \) \AA, as measured in regions B and E (see Figure
3), is exactly as predicted for a significant young stellar population but is
not expected for an old (15 Gyr) stellar population.

These results are consistent with TDS96, whose 1-d data did not allow spatial
resolution of the different regions presented here. Although they quote the
contributions of the various components as a fraction of a bin centered at 3620
\AA, it is possible to estimate the contribution this would give at the 5060
\AA~bin we used. The TDS96 data yield contributions of \( 71 \), \( 21 \)
and \( 8 \)\%, at \( 5060 \) \AA, of the continuum remaining after subtraction
of the nebular component for the 15 Gyr, 1 Gyr and power-law components respectively,
using their value of \( \alpha =-1.7 \). (Note that their young stellar population
is older than the one used in this paper.) Taking a broad spatial aperture corresponding
as closely as possible to the one used by TDS, containing regions D -- G, the
continuum fitting gives contributions of \( 60 \), \( 20 \), and \( 17.5 \)
\% for the 15 Gyr, 0.1 Gyr and power-law, which is entirely consistent given
that the young stellar components in the two sets of models used have different
ages, and also that our data provide much better constraints in the red.

It is possible to calculate the total mass of stars present in each component
from these results. The original \shortcite{Bruzual+C-1993} models are all normalized
to a total mass of \( 1\, \mbox {M}_{\odot } \), so it is easy to scale the
fluxes so as to give the contribution to the normalizing bin indicated by the
best fit results in Table \ref{tab: Continuum components}. The masses thus
derived are shown in Table \ref{tab: Masses}. The young stellar population
forms \( \sim 1\% \) of the total stellar mass in all the spatial regions,
and we believe this is the first published confirmation that the young stellar
populations in radio galaxies are spatially extended. Note that for the older,
1 Gyr population used by TDS, we would expect the young stellar population to
make a greater proportional contribution to the total mass.

Although we believe our models show strong evidence for a young stellar population
that is spatially extended across the entire nebulosity, we emphasize that the
models are not unique in terms of the age of the young stellar population. For
example, TDS96 found that a 1 Gyr starburst provided a good fit to the data.
More work is required to determine the age of the starburst accurately, although
we stress that the emission line flux measurements are not significantly affected
by the exact age of the starburst used in the continuum subtraction in the age
range \( 0.01-1 \) Gyr \cite{Dickson-1997}.

\subsection{Emission line kinematics}

\label{sec: Kinematics} Plots showing the emission-line kinematics are presented
in Figure \ref{fig: EL kinematics}. The bright {[}OIII{]}\( \lambda 5007 \)
line was measured in every spatial pixel, whereas the other lines were measured
in the regions identified above. Both velocity shifts and velocity widths of
the lines are plotted, with the widths quadratically corrected for instrumental
broadening using the measured widths of night sky lines as close as possible
in wavelength to the emission lines being studied. The `zero' of velocity shift
is defined as the wavelength of the line measured at the centre of \textbf{}the
\textbf{continuum} emission of the brighter (SE) nucleus, as determined by fitting
a Gaussian to its spatial profile. 

The plots are encouraging in terms of using 3C 321 to test the photoionization
models, since they suggest that, relative to other radio galaxies, 3C 321 is
kinematically undisturbed. Not only are the lines relatively narrow (FWHM \( <\sim 500 \)
kms\( ^{-1} \)) across much of the nebula, but there are no significant differences
in kinematics between lines of different ionization, as has been found in cases
of jet-cloud interactions \cite{Villar-Martin+T+M+etal-1999}. It is only in
the farthest spatial regions to the SE and NW that there is any sign of broadening
in the emission lines. In particular, region A (\( 7.5 \) arcsec NW of the
nucleus) shows evidence for significantly broader line profiles in all the measured
lines (FWHM \( \sim 500-600 \) kms\( ^{-1} \)).

The spatial variation of radial velocities across the nebula is also small,
with a spread of \( <200 \)kms\( ^{-1} \) in the range \( -6 \) to \( +2 \)
arcsec. The variation which is observed may be due as much to the structure
of the emission line nebula within the slit as to intrinsic velocity differences
between different regions, since the seeing (\( \sim 0.8 \) arcsec) was significantly
less than the slit width.

Generally, the values presented here correspond to typical narrow-line widths
in radio galaxies \cite{Baum+H+VB-1990}, and are entirely compatible with purely
gravitational motions \cite{Tadhunter+F+Q-1989}.

\subsection{Reddening}

\label{sec: Reddening} The line fluxes were corrected for reddening using the
\shortcite{Seaton-1979} re-normalization of the \shortcite{Nandy+T+J+M+W-1975}
extinction curve, with Case B recombination line values from \shortcite{Gaskell+F-1984}
for H\( \alpha  \)/H\( \beta  \) and from \shortcite{Osterbrock-1989} for
all other line ratios. The continuum-subtracted spectra allowed good fits to
be made to H\( \gamma  \) and H\( \delta  \), as well as the other brighter
members of the Balmer series, in all seven regions. The resulting E(B-V)values
are shown in Table \ref{tab: EL fluxes}. These values suggest no intrinsic
reddening in regions B \& D, and a maximum in region F. The high values in regions
E -- G are consistent with the HST image (Figure \ref{fig: HST Image}), which
shows a prominent dust lane across the SE edge of region F. These values of
E(B-V) are also comparable with the extinction measured in the dust lane of
Cygnus A of E(B-V) \( \sim 0.3-0.6 \) \cite{Tadhunter+M+R-1994}.

The estimated extinction values are smaller than the E(B-V) estimated by TDS96,
but it should be noted that their measurement of H\( \gamma  \)/H\( \beta  \)
was made before any continuum subtraction, and that therefore no correction
was made for underlying Balmer line absorption. Their measured value of H\( \gamma  \)/H\( \beta  \)
was therefore \( \sim 25\% \) lower than ours, which accounts for the apparent
discrepancy in E(B-V).

Despite their much larger value of E(B-V), TDS96 found that de-reddening the
continuum had a negligible effect on the continuum fitting, in terms of the
contributions of the different components. This suggests there is little to
be gained in repeating the continuum fitting/subtraction for our spectra after
correcting them for reddening. However, the emission line fluxes were themselves
corrected for intrinsic reddening, using the values shown in Table \ref{tab: EL fluxes}.

\subsection{Emission-line properties}

After subtracting the best-fit model continuum from the observed spectrum in
each region, the emission line fluxes in each spatial region were measured using
the Starlink package \noun{dipso} and corrected for the effects of intrinsic
reddening. The corrected fluxes relative to H\( \beta  \) are shown in Table
\ref{tab: Reddening corrected EL}. 

We can confirm the detection of a broad component to H\( \alpha  \) reported
by \shortcite{Filippenko-1987} in direct light. Whereas Filippenko only detected
the broad H\( \alpha  \) in the brighter nucleus (corresponding to apertures
E \& F), we also detect it in the fainter nucleus (aperture C), albeit at low
signal-to-noise. The FWHM velocity-widths of this component, where measured,
are shown in Table \ref{tab: EL fluxes}. These are consistent with the FWZI
\( >6000 \) kms\( ^{-1} \)measured by \shortcite{Filippenko-1987}. The equivalent
widths of the broad component, relative to the modelled power-law (i.e. quasar)
component of the continuum in the regions C, E, F and G are also shown in Table
\ref{tab: EL fluxes}. The fact that these are lower than the values found in
typical quasars (\( \sim 250 \) \AA, \shortcite{Espey+C+B+etal-1989}) suggests
that what we have modelled as the ``power law'' is not purely scattered quasar
light, but may include a contribution from a non-quasar source, such as stellar
populations with ages younger than those we have assumed in our fit.

Several features of these emission line data are worth highlighting:

\begin{itemize}
\item Apart from region A which shows a significantly lower ionization state, there
is a remarkable consistency of line ratios and ionization between every spatial
aperture. This lack of variation was first noted by \shortcite{Filippenko-1987} who
discussed the integrated spectra of the main NW and SE components and noted
their general similarity.
\item A high ionization state is indicated by unusually strong HeII\( \lambda 4686 \),
{[}NeV{]}\( \lambda 3426 \) in comparison with the nuclear NLR of most other
powerful radio galaxies (see Section \ref{sec: Photoionization models}, Figure
\ref{fig: Diagdiags}), and by the presence of the higher-ionization lines,
{[}FeVII{]}\( \lambda 6087 \) and {[}FeX{]}\( \lambda 6375 \), in regions
E, F and G.
\item Despite the high ionization state, the temperature-sensitive {[}OIII{]}\( \lambda 4363/\lambda 5007 \)
ratio is smaller than measured in the nuclear and extra-nuclear NLR of most
other powerful radio galaxies \cite{Tadhunter+R+M-1989,Binette+W+S-B-1996}. 
\end{itemize}

\subsection{Physical conditions and ionization parameter}

\label{sec: Phys conds}Assuming that the AGN is the dominant photoionizing
source, and that the line-emitting clouds are optically thick to the photoionizing
radiation, values for the ionization parameter, \( U \), (defined as the ratio
of the number density of photoionizing photons to the number density of photoionized
electrons) the electron density, \( n_{e} \), and the electron temperature,
\( T_{e} \), may be derived from the {[}OII{]}\( \lambda 3727 \)/{[}OIII{]}\( \lambda 5007 \),
{[}SII{]}\( \lambda 6716 \)/\( \lambda 6731 \) and {[}OIII{]}\( \lambda \lambda (5007+4959)/4363 \)
line ratios, respectively.

The temperature and density were calculated using the Lick 5-level atom calculator,
\noun{fivel}, \cite{DeRobertis+D+H-1987} \textbf{}which, given a nominal
density and value of the temperature diagnostic line ratio, can be used to give
the temperature of the line-emitting region, and vice versa. The results of
this process, using the de-reddened line ratios, are shown in Table \ref{tab: Physical conditions}.

\shortcite{Penston+R+etal-1990} give the following relationship between {[}OII{]}\( \lambda 3727 \),
{[}OIII{]}\( \lambda 5007 \), and \( U \): 

\begin{quotation}
{\par\centering \(  \log U=-2.74-\log \frac{\mbox{[OII]$\lambda$3727}}{\mbox{[OIII]$\lambda$5007}} \) \par}
\end{quotation}
$\frac{1}{2} $
This relationship was calculated using a set of models in which optically thick
clouds, with a constant gas density of 100 cm\( ^{-3} \) and solar abundances,
were photoionized by a central source. {[}OII{]}/{[}OIII{]} was chosen because
it is known to be insensitive to the detailed continuum shape, while retaining
sensitivity to \( U \) over a wide range of values.

Values of \( U \) were calculated for each region together with the density
and projected distance --- measured from the mean continuum centroid of the
SE nucleus --- and these results are shown in Table \ref{tab: Physical conditions}.

Again the results presented in Table \ref{tab: Physical conditions} show a
remarkable consistency in physical conditions and ionization between the various
regions.

\section{Testing the ionization models for 3C 321}

\label{sec: Discussion} The results presented so far emphasize the fact that
it is reasonable to assume that the ionization of the emission line gas in 3C
321 is dominated by AGN photoionization, with no clear evidence for shock-ionization.
The justifications for this are:

\begin{itemize}
\item \textbf{Kinematics.} The kinematics of the line emission gas are consistent
with purely gravitational motion, and imply that the gas is relatively undisturbed.
\item \textbf{Physical conditions.} Shock models which include a combination of cooling
and precursor gas (e.g., \citeN{Dopita+S-1996}), predict relatively high electron
temperatures (T \( >15,000 \)K), whereas the temperatures measured in several
apertures across 3C 321 are much lower than this.
\end{itemize}
Although there is no evidence for shock ionization, shocks may nonetheless influence
the physical conditions in 3C 321. Such influence may be indicated by the consistently
high densities measured across the central regions of 3C 321 (see Table \ref{tab: Physical conditions}).
Warm emission line clouds in pressure equilibrium with the hot X-ray haloes
of isolated elliptical galaxies and groups of galaxies would have electron densities
in the range \( n_{e}<10 \) cm\( ^{-3} \) at a similar radius \cite{Forman+J+T-1985,Ponman+A+J+etal-1994}\footnote{%
The X-ray luminosity of 3C 321 (\( <10^{43} \) erg s\( ^{-1} \)) is in the
range typical of isolated ellipticals or groups of galaxies \cite{Fabbiano+T+E+etal-1984}.
}. The larger densities measured in 3C 321 suggest a higher confining pressure,
perhaps associated with the cocoon of the radio source. Indeed, the cocoons
of radio sources are predicted to be substantially over-pressured with respect
to the ambient ISM. 

Alternatively, the warm photoionized clouds, initially over-pressured with respect
to their surroundings due the large temperature increase caused by photoionization,
may not yet have expanded to reach pressure equilibrium with the hot ISM.

\subsection{Photoionization models}

\label{sec: Photoionization models} Initially, simple photoionization models
comprising a power-law continuum, \( f_{\nu }\propto \nu ^{\alpha _{\nu }} \),
photoionizing an optically thick, solar-abundance medium, were used to predict
values of various emission line flux ratios, using the \noun{mappings} photoionization
code of \shortcite{Binette+W+R+etal-1997}. A selection of diagnostic diagrams
is shown in Figure \ref{fig: Diagdiags}. Two power-law sequences are plotted,
one with \( \alpha _{\nu } \) fixed at \( -1.5 \) and \( 10^{-4}<U<0.1 \),
and the second with \( U \) fixed at \( 0.01 \) with \( -2<\alpha _{\nu }<-1 \).
Note that an \( \alpha _{\nu }=-1.5 \) power-law continuum has been shown to
give good fits to the stronger emission lines in low redshift radio galaxies
\cite{Ferland+O-1986,Robinson+B+F+T-1987}. Also plotted for comparison are
the emission line ratios for the nuclear NLR of various powerful radio galaxies
from the literature. While these simple power-law photoionization models are
in reasonable agreement with the line ratios measured in 3C 321 on some of the
diagnostic diagrams, there are several problems. These include the following.

\begin{itemize}
\item There is a general lack of consistency between the various diagnostic diagrams,
in the sense that, even though the points may fall close to the power-law U
sequence on some of the diagrams, the points move to different positions along
the U sequence between the different diagrams.
\item {[}NII{]}\( \lambda 6583 \)/{[}OII{]}\( \lambda \lambda 3727 \) is under-predicted
by the power-law models (Figure \ref{fig: Diagdiags}c). This problem has been
noted before in the case of Cygnus A \cite{Tadhunter+M+R-1994} where it was
ascribed to an N/O abundance ratio which is enhanced relative to the solar value.
On the basis of Figure \ref{fig: Diagdiags}c it seems safe to conclude that
the N/O abundance ratio is also enhanced in 3C321 and several other powerful
radio galaxies, given that the {[}NII{]}\( \lambda 6583 \)/{[}OII{]}\( \lambda 3727 \)
ratio is relatively insensitive to the ionizing continuum shape, ionization
parameter and cloud column density.
\item HeII\( \lambda 4686 \)/H\( \beta  \) is under-predicted by the power-law sequence
(see Figures \ref{fig: Diagdiags}a, b \& e). This is a particular problem,
since the ratio is not sensitive to abundances, physical conditions or ionization
parameter.
\item {[}OI{]}\( \lambda 6300 \)/{[}OIII{]}\( \lambda 5007 \) and {[}NeIII{]}\( \lambda 3869 \)/{[}NeV{]}\( \lambda 3426 \)
are also not well-fitted by the power-law sequence (see Figures \ref{fig: Diagdiags}e,
f \& g), falling at or off the extreme high ionization end of the U sequence.
\item The locus of points formed by 3C321 and the other radio galaxies in HeII\( \lambda 4686 \)/H\( \beta  \)
vs \\{[}NeIII{]}\( \lambda 3869 \)/{[}NeV{]}\( \lambda 3426 \) diagram (Figure
\ref{fig: Diagdiags}e) is almost orthogonal to the power-law U-sequence, with
3C321 at the extreme high ionization end of the sequence.
\end{itemize}
Next, in an attempt to get a better fit to the high ionization lines, photoionization
models were constructed for optically thick clouds photoionized by a black body
spectrum with a range of characteristic temperatures and ionization parameters.
Again two different sets of models are shown in Figure \ref{fig: Diagdiags}.
For one set, a black-body temperature of \( T_{b}=160,000 \) K was assumed,
and the ionization parameter was varied over the range \( 10^{-4}<U<0.1 \),
while for the other set \( U \) was fixed at \( 10^{-2} \) and the temperature
varied in the range \( 60,000\mbox {\, K}<T_{b}<200,000\mbox {\, K} \). Although
the \( T_{b}=160,000 \) K sequence gives better fits to the observed HeII\( \lambda 4686 \)/H\( \beta  \)
and {[}OI{]}\( \lambda 6300 \)/{[}OIII{]}\( \lambda 5007 \) ratios, the {[}NeIII{]}\( \lambda 3869 \)/{[}NeV{]}\( \lambda 3426 \)
ratio is still a problem and, as with the power law models, there is a general
lack of consistency between diagrams in the position of the points along the
\( U \) sequence.

We conclude that the optically thick models cannot fit the ratios of all the
emission lines in a consistent fashion between the different diagnostic diagrams,
regardless of the shape of the ionizing continuum. In the light of this conclusion,
we consider a third set of the models which include a mixture of optically thick
and optically thin clouds. Developed by \shortcite{Binette+W+S-B-1996} in order
to explain the discrepant high ionization line ratios and high electron temperatures
observed in many active galaxies, these models are shown in Figure \ref{fig: Diagdiags}
as sequences in \( A_{M/I} \), the ratio of the solid angle covered by the
optically thin component relative to that covered by the optically thick component.
Covering the range \( 0.04<A_{M/I}<16 \), they were generated using the tabulated
values from \shortcite{Binette+W+S-B-1996} for an illuminating power law with
\( \alpha _{\nu }=-1.3 \). As can be seen from Figure \ref{fig: Diagdiags},
the mixed medium models provide a remarkably good fit to the observed line ratios
on most of the diagnostic diagrams, with a consistent position of the points
along the \( A_{M/I} \) sequence in all diagrams. Particularly notable is the
fact that the \( A_{M/I} \) sequences provide an excellent fit to the locus
of radio galaxy points on the {[}NeIII{]}\( \lambda 3869 \)/{[}NeV{]}\( \lambda 3426 \)
vs HeII/H\( \beta  \) plot (Figure \ref{fig: Diagdiags}e), which is problematic
for the optically thick models. In terms of the diagnostic diagrams, the only
significant problem with the mixed medium models is that they over-predict the
temperature-sensitive {[}OIII{]}\( \lambda 4363/\lambda 5007 \) ratio. However,
better agreement with this line ratio can be obtained with super-solar abundances
(see Figs. 12 \& 13 of \shortcite{Binette+W+S-B-1996}). 

A general problem for all photoionization models is the observed uniformity
of the line ratios across the emission line nebula. In the case of the optically
thick models, such uniformity would imply a small range in ionization parameter
which in turn would require a \( 1/r^{2} \) dependence in the density, where
\( r \) is the distance from the illuminating source. However, such a density
dependence seems unlikely, given that the density, where it can be measured,
remains constant over a range of projected radii, and also given that one of
the emission line regions (C) is associated with the secondary nucleus. 

The uniformity may be even more of a problem for the mixed medium models, since
it requires not only a small variation in \( A_{M/I} \) between the different
regions, but also consistent column depths for the optically thick and thin
components and consistent properties for the photoionizing continuum. In the
context of the mixed medium models, such uniformity may imply some as yet unrecognized
physical mechanism which maintains a ``natural'' \( A_{M/I} \) ratio over
a range of radii.

\subsection{Photoionizing photon luminosity, \emph{Q}}

Another way of testing the photoionization models is to calculate the value
of \( Q \), the flux of ionizing photons per unit solid angle and check for
consistency with other measures of the AGN brightness. \( Q \) may be calculated
from the various physical quantities estimated in Section \ref{sec: Phys conds}
according to the formula \( Q=cn_{e}Ud^{2} \), where \( d \) is the (projected)
distance of the line-emitting region from the source \cite{Robinson+B+F+T-1987}.

As in the previous section, we begin by considering results derived from photoionization
models for optically thick clouds. Table \ref{tab: Physical conditions} shows
values of \( Q \) calculated for each region, under the assumption that the
projected distance measured from our spectra correspond to the real AGN-cloud
distances. The results reveal two problems associated with the values of \( Q \)
calculated for these optically thick models:

\begin{enumerate}
\item \textbf{Absolute value.} \( Q \) is much higher than in other low-redshift
radio-loud objects -- up to 2 orders of magnitude greater than the value measured
for Cygnus A (\( Q\sim 10^{54}\mbox {photon\, s}^{-1}\mbox {sr}^{-1} \) \cite{Tadhunter+M+R-1994}),
and \( \sim 3 \) times greater than the most powerful radio-loud quasar in
the local universe, 3C 273 (\( Q\sim 3\times 10^{55}\mbox {photon\, s}^{-1}\mbox {sr}^{-1} \)
(based upon a linear interpolation across the EUV gap \cite{Elvis+W+M+etal-1994}.)
This is surprising in the context of the correlation between optical and radio
properties found by \shortcite{Rawlings+S-1991} since 3C 321 is not an unusually
luminous radio source -- \( P_{178MHz} \) is roughly \( 200 \) times less
than for Cygnus A \cite{Spinrad+M+A+D-1985}. Note that the values of \( Q \)
in Table \ref{tab: Physical conditions} are lower limits, since the distances
used in the calculation are projected rather than true distances. Region C gives
the most reliable estimate of \( Q \) because it is the region with the least
fractional uncertainty in distance from the putative illuminating source; it
also gives the largest value of \( Q \).
\item \textbf{Predicted far-infrared flux.} Using the value of \( Q \) it is possible
to predict the far-infrared luminosity, assuming that all the photoionizing
photons incident on the dusty torus are reprocessed into the far infrared. If
we assume a cone opening half-angle of \( 50^{\circ } \), the solid angle subtended
by the torus is \( 4\pi \cos 50=8.1\mbox {\, sr} \). Assuming an isotropic
illuminating continuum with the form of a \( 160,000 \) K black body, as indicated
by the photoionization models above, with a mean photon energy of \( \varepsilon =42 \)
eV, then the total bolometric luminosity expected in the far infrared from dust-reprocessing
is \( L_{IR}=Q\times \varepsilon \times 4\pi \cos 50=5\times 10^{39}\mbox {\, W} \). \shortcite{Golombek+M+N-1988}
measure the observed \( 25\, \mu  \)m and \( 60\, \mu  \)m fluxes of 3C 321
as \( 353 \) and \( 1067 \) mJy respectively. Assuming that a power-law obtains
between \( 25 \) and \( 60\, \mu  \)m, the integrated rest-frame luminosity
between these wavelengths is \( L_{25-60}=2\times 10^{38}\mbox {\, W} \) --
a factor of \( \sim 25 \) less than that expected from the measured value of
\( Q \).
\end{enumerate}
By contrast, if the mixed medium models are correct, \shortcite{Binette+W+S-B-1996}
estimate that the `true' ionization parameter in the optically thick clouds,
from which almost all of the {[}SII{]} density diagnostic emission originates,
is \( U\sim 5\times 10^{-4} \). The last column of Table \ref{tab: Physical conditions}
shows \( Q \) for the different regions calculated using the value of \( U \)
appropriate for a mixed medium model. Repeating the above calculation for the
expected re-processed far infrared with the values of \( Q \) calculated for
a mixed medium gives \( L_{IR}\sim 2.5\times 10^{38} \), which is close to
the value inferred from the observations of \shortcite{Golombek+M+N-1988}. Although
there is some uncertainty about the exact value of \( U \) for the mixed medium
model, the estimates of \( Q \) based on this model are clearly more realistic
than those derived for the optically thick case, although region C, with the
most reliable estimate of \( Q \), still gives a value \( \sim 5 \) times
greater than that estimated for Cygnus A \cite{Tadhunter+M+R-1994}. This is
perhaps not too surprising since it has been shown that there is a large degree
of scatter in quasar power for a given extended radio power \cite{Tadhunter+M+R+D+V-M+F-1998}.
Also, it has been suggested that Cygnus A has enhanced radio emission in relation
to its emission line luminosity, probably as a result of its occupying a dense
cluster environment \cite{Barthel+A-1996}. By comparison, 3C 321 is an isolated
galaxy.

The more realistic estimates of \( Q \) for the mixed medium models further
strengthen the case of these models in comparison with the models which assume
optically thick clouds.

\subsection{The `Q effect'}

\label{sec: Variation in Q}Whether optically thick or mixed medium models are
assumed for the calculation of \( Q \), the resulting values show a prominent
dip across the object, which coincides with the assumed position of the continuum
source. As \( Q \) is the ionizing luminosity `seen' by the clouds, it should
be constant for all positions, since all clouds should `see' the central source.
This `Q effect' has been reported before in the case of Seyfert galaxies \cite{Metz-1997,Metz+A+R+T-1997}.
These authors noted that the size of the region across which the dip in \( Q \)
was most acute (corresponding with our region F) was on the same scale as the
seeing disc. This led them to infer the presence of an unresolved dense core,
or inner narrow line region (INLR), situated between the conventional broad
and narrow line regions. The HST {[}OIII{]} image (Figure \ref{fig: HST Image})
provides some evidence that this may also be the case in 3C 321, with the brightest
pixels bordering the dust-lane in the SE nucleus. However, the consistency of
our emission line data across the object show no evidence for such a dense inner
region, and the highest electron densities are found in the regions furthest
from the centre.

A second, more plausible effect is suggested by our data -- owing to the consistency
of the emission lines across the nebula, most of the variation in \( Q \) is
due to the increase in measured distance from the continuum centre (Table \ref{tab: Physical conditions}).
These distances are clearly projected distances, and therefore represent lower
limits to the true values. There may therefore be a projection effect at work,
whereby the line-emitting clouds in regions E \& F are in fact some distance
away from the AGN along our line of sight. With respect to the unified schemes
of radio galaxies and quasars \cite{Barthel-1989,Lawrence-1991}, this would
imply that we must be viewing the central regions of 3C 321 close to the edges
of the photoionized cones (any further towards the centres of the cones would
result in a direct view of the quasar nucleus). In this context, it interesting
to note that infrared observations by \shortcite{Tadhunter+P+A+etal-1999} show
that the emission cones in Cygnus A are hollowed out, with much of the line
emission generated at the edges of the cones. This suggests that the the projection
effects indicated by the dip in \( Q \) may plausibly occur without exposing
the hidden quasar nucleus. The current best estimates of cone opening angles,
derived from examination of the statistics of AGN \cite{Lawrence-1991}, therefore
imply inclinations of the radio axis with respect to the plane of the sky in
the range \( \sim 30-45^{\circ } \) if the edge of the cone in 3C 321 is close
to the line of sight. In this case we would expect the core radio flux to be
boosted with respect to that of the lobes, as is the case for broad line radio
galaxies (BLRGs). \shortcite{Morganti+O+R+T+M-1997} derive median values of
of \( R_{2.3GHz}=0.0269 \) and \( 0.00295 \) for broad- and narrow-line radio
galaxies respectively, where \( R_{2.3GHz}\equiv P_{2.3GHz,core}/(P_{2.3GHz,total}-P_{2.3GHz,core}) \).
The fact that \( R_{5GHz}=0.0238 \) for 3C 321 \cite{Nilsson-1998} suggests
that these projection effects may indeed be important, as this is clearly closer
to the median BLRG value than that of NLRGs. This is also supported by the one-sided
core-jet structure to the NW, visible in the maps of \shortcite{Baum+H+B+vB+M-1988}. 

In summary, it is likely that projection effects can account for the dip in
\( Q \) at the SE nucleus and that the true value of \( Q \) is close to that
measured in region C, the furthest spatial bin from the nucleus in which we
can measure \( Q \).

\section{Conclusion}

We have presented high quality, long-slit spectral data for 3C 321. The kinematics,
emission-line ratios and physical properties show an unusually small variation
across the spatial extent of the object, and provide good evidence that 3C 321
is mainly photoionized, with little or no contribution from shock ionization.
3C 321 is therefore an excellent source to use to test the photoionization models
in depth. 

We have carried out accurate continuum modelling and subtraction to allow the
measurement of faint emission lines so as to better constrain the photoionization
models, and allow calculation of physical conditions across the nebula. The
continuum modelling requires the presence of a significant, spatially extended,
young stellar population to give good fits to the data.

Comparison of emission-line flux ratios with the photoionization models shows
that the mixed medium models -- involving a combination of clouds which are
optically thick and thin to the ionizing radiation -- provide an excellent fit
to the observed line ratios, especially if super-solar abundances are assumed.
Conventional models, which assume a single population of optically thick clouds,
do not provide a good fit to the observed line ratios, and require differing
values for parameters such as \( U \) to reproduce the line ratios in different
diagnostic diagrams. 

Our results also show that caution is required when using the emission line
properties to calculate the ionizing luminosity of the illuminating AGN (\( Q \)).
Under the assumption that the line emitting clouds are optically thick, we find
that the estimated value of \( Q \) is unrealistically large, given the radio
and far-infrared properties of the source. However, the lower value of ionization
parameter appropriate for the mixed medium model leads to an estimate of \( Q \)
which is more consistent with the estimated far-infrared luminosity of 3C 321.

The unusual degree of consistency of the emission line ratios across the nebula
presents difficulties for any ionization model due to the need for consistent
physical conditions and ionization over large range of distances from the illuminating
nucleus. This is a particular problem for the mixed medium model which provides
the best fit to the line ratios, since the uniformity requires the ratio of
covering factors of the optically thick and thin components to remain remarkably
constant across the nebula.\\

\textbf{Acknowledgements.} The authors acknowledge the data analysis facilities
provided by the Starlink Project which is run by CCLRC on behalf of PPARC. In
addition, the following Starlink software packages have been used: FIGARO, KAPPA,
\& DIPSO. Luc Binette provided useful discussions regarding the details of the
mixed medium models. AR would like to thank the Royal Society for financial
support, and TGR acknowledges a PPARC studentship.

The WHT is operated on the island of La Palma by the Isaac Newton Group in the
Spanish Observatorio del Roque de los Muchachos of the Instituto de Astrofisica
de Canarias. Figure \ref{fig: HST Image} is based on observations made with
the NASA/ESA Hubble Space Telescope, obtained from the data archive at the Space
Telescope Science Institute. STScI is operated by the Association of Universities
for Research in Astronomy, Inc. under NASA contract NAS 5-26555.

\bibliographystyle{paper}
\bibliography{paper}

\begin{table}[p]
{\centering \begin{tabular}{|c|c|c|c|c|c|}
\hline 
Region&
15 Gyr flux\( ^{\dagger } \)&
Power law flux\( ^{\dagger } \)&
\( \alpha  \)&
\( \chi _{min,red}^{2} \)&
Goodness-of-fit\( ^{\ddagger } \)\\
\hline 
\hline 
A&
\( 63_{-6}^{+10} \)&
\( 27_{-9}^{+6} \)&
\( -0.5_{-1.0}^{+0.4} \)&
\( 10.98 \)&
\( 0.000 \)\\
B&
\( 67_{-8}^{+6} \)&
\( 31_{-6}^{+8} \)&
\( -1.3\pm 0.6 \)&
\( 2.45 \)&
\( 0.000 \)\\
C&
\( 80_{-11}^{+7} \)&
\( 22_{-7}^{+10} \)&
\( -0.9_{-1.0}^{+0.8} \)&
\( 0.77 \)&
\( 0.820 \)\\
D&
\( 63_{-4}^{+6} \)&
\( 23_{-5}^{+4} \)&
\( -1.3_{-0.6}^{+0.4} \)&
\( 4.09 \)&
\( 0.000 \)\\
E&
\( 69\pm 7 \)&
\( 27^{+8}_{-6} \)&
\( -1.1_{-0.8}^{+0.6} \)&
\( 1.44 \)&
\( 0.051 \)\\
F&
\( 71_{-8}^{+10} \)&
\( 30_{-8}^{+7} \)&
\( 0.1_{-0.6}^{+0.4} \)&
\( 0.55 \)&
\( 0.982 \)\\
G&
\( 78_{-7}^{+6} \)&
\( 20_{-6}^{+7} \)&
\( -0.9^{+0.6}_{-0.8} \)&
\( 1.96 \)&
\( 0.001 \)\\
\hline 
\end{tabular}\par}

\caption{\label{tab: 2-component} 2-component fits to the continuum in the each of
the regions A-G. Errors quoted denote the range of values which give \protect\( \chi ^{2}\protect \)
within the range \protect\( \chi _{min}^{2}<\chi ^{2}<\chi _{min}^{2}+\Delta (\chi ^{2})\protect \),
where \protect\( \Delta \chi ^{2}\protect \) is the interval in \protect\( \chi ^{2}\protect \)
corresponding to the 68\% confidence level \protect\cite{Press+T+V+F-1992}. The value
of \protect\( \chi _{min,red}^{2}\protect \) measured for A is almost certainly
an overestimate -- the error on each data point is much greater than the flux
calibration error due to noise in the continuum.}

\( ^{\dagger } \)As a percentage of the measured continuum after subtraction
of the nebular component in the rest wavelength region \( 5046.0-5075.0 \)
\AA.

\( ^{\ddagger } \)Defined as the formal statistical probability that this value
of \( \chi _{min,red}^{2} \) has not been arrived at by chance.
\end{table}

\begin{table}[p]
{\centering \begin{tabular}{|c|c|c|c|c|c|c|}
\hline 
Region&
15 Gyr flux\( ^{\dagger } \)&
0.1 Gyr flux\( ^{\dagger } \)&
Power law flux\( ^{\dagger } \)&
\( \alpha  \)&
\( \chi _{min,red}^{2} \)&
Goodness-of-fit\\
\hline 
\hline 
A&
\( 62.5_{-12.5}^{+10.0} \)&
\( 0_{-0}^{+5} \)&
\( 27.5\pm 10.0 \)&
\( -0.5_{-0.8}^{+0.6} \)&
\( 11.32 \)&
\( 0.000 \)\\
B&
\( 57.5\pm 12.5 \)&
\( 20.0_{-7.5}^{+12.5} \)&
\( 22.5_{-17.5}^{+10.0} \)&
\( -0.9_{-2.4}^{+1.0} \)&
\( 1.24 \)&
\( 0.139 \)\\
C&
\( 72.5_{-17.5}^{+12.5} \)&
\( 10.0_{-10.0}^{+12.5} \)&
\( 20.0\pm 12.5 \)&
\( -0.3_{-1.8}^{+1.0} \)&
\( 0.64 \)&
\( 0.921 \)\\
D&
\( 52.5\pm 5.0 \)&
\( 40.0_{-7.5}^{+5.0} \)&
\( 0.0_{-0.0}^{+7.5} \)&
n/a&
\( 0.74 \)&
\( 0.818 \)\\
E&
\( 55.0_{-10.0}^{+17.5} \)&
\( 25.0\pm 12.5 \)&
\( 17.5_{-17.5}^{+7.5} \)&
\( -0.1_{-5.0}^{+1.0} \)&
\( 0.38 \)&
\( 0.999 \)\\
F&
\( 60.0_{-7.5}^{+17.5} \)&
\( 15.0_{-15.0}^{+7.5} \)&
\( 25.0_{-7.5}^{+12.5} \)&
\( 0.9_{-1.2}^{+0.6} \)&
\( 0.44 \)&
\( 0.995 \)\\
G&
\( 67.5_{-12.5}^{+10.0} \)&
\( 20_{-7.5}^{+12.5} \)&
\( 12.5_{-10.0}^{+7.5} \)&
\( 0.1_{-1.6}^{+2.0} \)&
\( 1.12 \)&
\( 0.253 \)\\
\hline 
\end{tabular}\par}

\caption{\label{tab: Continuum components} As Table \ref{tab: 2-component}, but for
three components. The power law flux in region D makes no contribution to the
best fit model and so all values of \protect\( \alpha \protect \) give an equally
good fit.}

\( ^{\dagger } \)As a percentage of the measured continuum in the rest wavelength
bin \( 5046.0-5075.0 \) \AA.
\end{table}

\begin{table}[p]
{\centering \begin{tabular}{|c|c|c|c|}
\hline 
Region&
Mass in 15 Gyr component &
Mass in 0.1 Gyr component&
Percentage of young \\
&
\( /10^{10}\mbox {M}_{\odot } \)&
\( /10^{8}\mbox {M}_{\odot } \)&
 stars by mass\\
\hline 
\hline 
A&
\( 0.23 \)&
---&
---\\
B&
\( 0.75 \)&
\( 0.3 \)&
\( 0.4 \)\\
C&
\( 4.19 \)&
\( 0.6 \)&
\( 0.1 \)\\
D&
\( 3.38 \)&
\( 2.6 \)&
\( 0.8 \)\\
E&
\( 3.79 \)&
\( 1.7 \)&
\( 0.5 \)\\
F&
\( 4.98 \)&
\( 1.3 \)&
\( 0.3 \)\\
G&
\( 2.52 \)&
\( 0.8 \)&
\( 0.3 \)\\
\hline 
\hline 
Total&
\( 19.9 \)&
\( 7.2 \)&
\( 0.4 \)\\
\hline 
\end{tabular}\par}

\caption{\label{tab: Masses} Total mass of stars in each component. These values are
obtained by scaling the mass of the original models (1 \protect\( \mbox {M}_{\odot }\protect \))
to give the luminosity (\protect\( L_{\odot }/\protect \)\AA) in the normalizing
bin indicated by the model fits.}
\end{table}

\begin{table}[p]
{\centering \begin{tabular}{|l|ccccccc|}
\hline 
\textbf{Probable}&
\multicolumn{7}{|c|}{Measured values relative to H\( \beta  \) }\\
\textbf{ID}&
A&
B&
C&
D&
E&
F&
G\\
\hline 
\hline 
{[}NeV{]}\( \lambda 3346 \)&
---&
\( 36\pm 2 \)&
\( 27\pm 1 \)&
\( 24\pm 2 \)&
\( 46\pm 3 \)&
\( 37\pm 1 \)&
\( 35\pm 1 \)\\
{[}NeV{]}\( \lambda 3426 \)&
\( 50\pm 10 \)&
\( 100\pm 6 \)&
\( 75\pm 2 \)&
\( 65\pm 6 \)&
\( 127\pm 9 \)&
\( 104\pm 3 \)&
\( 97\pm 3 \)\\
{[}OII{]}\( \lambda 3727^{a} \)&
\( 229\pm 12 \)&
\( 131\pm 4 \)&
\( 128\pm 2 \)&
\( 118\pm 5 \)&
\( 141\pm 6 \)&
\( 111\pm 2 \)&
\( 98\pm 2 \)\\
{[}NeIII{]}\( \lambda 3869 \)&
\( 77\pm 9 \)&
\( 77\pm 4 \)&
\( 69\pm 2 \)&
\( 55\pm 7 \)&
\( 88\pm 8 \)&
\( 68\pm 2 \)&
\( 62\pm 2 \)\\
{[}NeIII{]}\( \lambda 3964^{b} \)&
\( 24\pm 6 \)&
\( 32\pm 3 \)&
\( 30\pm 2 \)&
\( 45\pm 6 \)&
\( 24\pm 7 \)&
\( 32\pm 2 \)&
\( 32\pm 2 \)\\
{[}SII{]}\( \lambda 4069^{a} \)&
\( 12\pm 5 \)&
\( 6\pm 2 \)&
\( 5\pm 1 \)&
---&
---&
\( 4\pm 1 \)&
\( 4.3\pm 1.0 \)\\
H\( \delta  \)&
\( 26\pm 5 \)&
\( 26\pm 3 \)&
\( 14\pm 1 \)&
\( 18\pm 3 \)&
\( 13\pm 3 \)&
\( 18\pm 1 \)&
\( 17\pm 1 \)\\
H\( \gamma \lambda 4340 \)&
\( 34\pm 5 \)&
\( 43\pm 3 \)&
\( 40\pm 1 \)&
\( 51\pm 5 \)&
\( 37\pm 4 \)&
\( 36\pm 2 \)&
\( 39\pm 4 \)\\
{[}OIII{]}\( \lambda 4363 \)&
---&
\( 10\pm 3 \)&
\( 7\pm 1 \)&
\( 5.3\pm 3.2 \)&
\( 9\pm 3 \)&
\( 8\pm 1 \)&
\( 13\pm 3 \)\\
HeII\emph{\( \lambda 4686 \)}&
---&
\( 50\pm 3 \)&
\( 51\pm 1 \)&
\( 47\pm 4 \)&
\( 54\pm 3 \)&
\( 45\pm 1 \)&
\( 47\pm 1 \)\\
({[}ArIV{]})\( \lambda 4713 \)&
---&
\( 6\pm 2 \)&
\( 8\pm 2 \)&
---&
---&
\( 9\pm 1 \)&
\( 8\pm 2 \)\\
H\( \beta \lambda 4861 \)&
\( 100 \)&
\( 100 \)&
\( 100 \)&
\( 100 \)&
\( 100 \)&
\( 100 \)&
\( 100 \)\\
{[}OIII{]}\( \lambda 4959 \)&
\( 288\pm 14 \)&
\( 305\pm 7 \)&
\( 288\pm 4 \)&
\( 301\pm 10 \)&
\( 366\pm 13 \)&
\( 356\pm 6 \)&
\( 363\pm 6 \)\\
{[}OIII{]}\( \lambda 5007 \)&
\( 860\pm 41 \)&
\( 910\pm 21 \)&
\( 902\pm 11 \)&
\( 899\pm 30 \)&
\( 1094\pm 40 \)&
\( 1062\pm 17 \)&
\( 1083\pm 19 \)\\
{[}NI{]}\( \lambda 5199 \)&
---&
---&
\( 5\pm 1 \)&
\( <6 \)&
---&
\( 6\pm 1 \)&
\( 5\pm 1 \)\\
{[}FeVII{]}\( \lambda 6087 \)&
---&
---&
---&
---&
---&
\( 22\pm 2 \)&
\( 13\pm 1 \)\\
{[}OI{]}\( \lambda 6300 \)&
---&
\( 10\pm 2 \)&
\( 17\pm 3 \)&
---&
\( 17\pm 5 \)&
\( 18\pm 1 \)&
\( 18\pm 1 \)\\
{[}OI{]}\( \lambda 6364 \)&
---&
\( <4 \)&
\( 6\pm 1 \)&
---&
---&
\( 6\pm 1 \)&
\( 6\pm 1 \)\\
{[}FeX{]}\( \lambda 6375 \)&
---&
---&
---&
---&
\( 8\pm 3 \)&
\( 7\pm 1 \)&
\( 4\pm 2 \)\\
{[}NII{]}\( \lambda 6548 \)&
\( 54\pm 3 \)&
\( 47\pm 1 \)&
\( 70\pm 2 \)&
\( 52\pm 2 \)&
\( 77\pm 3 \)&
\( 82\pm 2 \)&
\( 57\pm 3 \)\\
H\( \alpha \lambda 6563 \)N&
\( 406\pm 22 \)&
\( 321\pm 8 \)&
\( 381\pm 6 \)&
\( 283\pm 10 \)&
\( 398\pm 15 \)&
\( 483\pm 8 \)&
\( 434\pm 11 \)\\
H\( \alpha \lambda 6563 \)B&
---&
---&
\( 35\pm 14 \)&
---&
\( 88\pm 14 \)&
\( 115\pm 16 \)&
\( 91\pm 24 \)\\
{[}NII{]}\( \lambda 6583 \)&
\( 163\pm 10 \)&
\( 140\pm 4 \)&
\( 212\pm 5 \)&
\( 155\pm 7 \)&
\( 231\pm 9 \)&
\( 247\pm 6 \)&
\( 170\pm 8 \)\\
HeI\( \lambda 6679 \)&
---&
\( <2 \)&
\( 8\pm 3 \)&
\( <2 \)&
---&
\( 7\pm 1 \)&
\( 6\pm 2 \)\\
{[}SII{]}\( \lambda 6716 \)&
\( 88\pm 13 \)&
\( 46\pm 2 \)&
\( 60\pm 1 \)&
\( 40\pm 3 \)&
\( 63\pm 3 \)&
\( 75\pm 1 \)&
\( 58\pm 1 \)\\
{[}SII{]}\( \lambda 6731 \)&
\( 69\pm 9 \)&
\( 31\pm 2 \)&
\( 57\pm 1 \)&
\( 40\pm 3 \)&
\( 57\pm 3 \)&
\( 63\pm 1 \)&
\( 50\pm 1 \)\\
{[}ArIII{]}\( \lambda 7136 \)&
---&
\( 16\pm 2 \)&
\( 24\pm 2 \)&
\( 19\pm 3 \)&
\( 39\pm 7 \)&
\( 33\pm 2 \)&
\( 31\pm 2 \)\\
\hline 
\hline 
H\( \beta  \) flux (\( 10^{-15} \)&
\( 0.24\pm 0.01 \)&
\( 0.51\pm 0.01 \)&
\( 1.63\pm 0.02 \)&
\( 0.56\pm 0.02 \)&
\( 0.53\pm 0.02 \)&
\( 1.54\pm 0.02 \)&
\( 1.23\pm 0.02 \)\\
ergs\( ^{-1} \)cm\( ^{-2} \)\AA\( ^{-1} \))&
&
&
&
&
&
&
\\
\hline 
H\( \beta  \) equivalent&
\( 29\pm 7 \)&
\( 15\pm 1 \)&
\( 11.1\pm 0.2 \)&
\( 3.6\pm 0.1 \)&
\( 2.6\pm 0.1 \)&
\( 7.6\pm 0.2 \)&
\( 13.0\pm 0.4 \)\\
 width (\AA)&
&
&
&
&
&
&
\\
\hline 
Broad H\( \alpha  \) &
---&
---&
\( 4000\pm 1000 \)&
---&
\( 3300\pm 400 \)&
\( 4300\pm 800 \)&
\( 3400\pm 800 \)\\
FWHM (kms\( ^{-1} \)).&
&
&
&
&
&
&
\\
\hline 
Broad H\( \alpha  \)&
&
&
&
&
&
&
\\
equiv. width&
---&
---&
\( 46\pm 34 \)&
---&
\( 28\pm 20 \)&
\( 59\pm 25 \)&
\( 200\pm 150 \)\\
 (\AA).&
&
&
&
&
&
&
\\
\hline 
E(B-V) \( ^{c} \)&
\( 0.21\pm 0.02 \)&
\( 0 \)&
\( 0.31\pm 0.01 \)&
\( 0 \)&
\( 0.38\pm 0.04 \)&
\( 0.50\pm 0.01 \)&
\( 0.41\pm 0.01 \)\\
\hline 
\end{tabular}\par}

\caption{\label{tab: EL fluxes} Measured emission line fluxes as a percentage of flux
in H\protect\( \beta \protect \). The narrow and broad components of H\protect\( \alpha \protect \)
are denoted by N and B respectively. All lines in the wavelength ranges covered
by the reduced spectra are shown. All quoted errors are Gaussian-fitting errors,
and do not include a contribution from flux calibration errors (estimatd as
\protect\( 5\%\protect \)). Also shown are the equivalent widths of H\protect\( \beta \protect \)
relative to the total continuum and of broad H\protect\( \alpha \protect \)
relative to the power-law component of the continuum model in each region.}

\( ^{a} \)Values for blends.

\( ^{b} \)Includes contribution from H\( \epsilon  \)

\( ^{c} \)Weighted means deduced from H\( \alpha  \)/H\( \beta  \), H\( \gamma  \)/H\( \beta  \)
and H\( \delta  \)/H\( \beta  \) and assuming Case B.
\end{table}

\begin{table}[p]
{\centering \begin{tabular}{|l|ccccccc|}
\hline 
\textbf{Probable}&
\multicolumn{7}{|c|}{Reddening corrected values relative to H\( \beta  \) }\\
\textbf{ID}&
A&
B&
C&
D&
E&
F&
G\\
\hline 
\hline 
{[}NeV{]}\( \lambda 3346 \)&
---&
\( 36\pm 2 \)&
\( 40\pm 1 \)&
\( 24\pm 2 \)&
\( 74\pm 5 \)&
\( 70\pm 2 \)&
\( 59\pm 2 \)\\
{[}NeV{]}\( \lambda 3426 \)&
\( 64\pm 13 \)&
\( 100\pm 6 \)&
\( 107\pm 3 \)&
\( 65\pm 6 \)&
\( 197\pm 13 \)&
\( 185\pm 6 \)&
\( 157\pm 5 \)\\
{[}OII{]}\( \lambda 3727^{a} \)&
\( 274\pm 15 \)&
\( 131\pm 4 \)&
\( 166\pm 3 \)&
\( 118\pm 5 \)&
\( 194\pm 9 \)&
\( 172\pm 4 \)&
\( 141\pm 3 \)\\
{[}NeIII{]}\( \lambda 3869 \)&
\( 90\pm 11 \)&
\( 77\pm 4 \)&
\( 86\pm 2 \)&
\( 55\pm 7 \)&
\( 117\pm 10 \)&
\( 99\pm 3 \)&
\( 85\pm 2 \)\\
{[}NeIII{]}\( \lambda 3964^{b} \)&
\( 28\pm 6 \)&
\( 32\pm 3 \)&
\( 37\pm 2 \)&
\( 45\pm 6 \)&
\( 31\pm 9 \)&
\( 45\pm 3 \)&
\( 42\pm 2 \)\\
{[}SII{]}\( \lambda 4069^{a} \)&
\( 14\pm 6 \)&
\( 6\pm 2 \)&
\( 6\pm 1 \)&
---&
---&
\( 5\pm 1 \)&
\( 6\pm 1 \)\\
H\( \delta  \)&
\( 29\pm 5 \)&
\( 26\pm 3 \)&
\( 17\pm 1 \)&
\( 18\pm 3 \)&
\( 16\pm 4 \)&
\( 24\pm 1 \)&
\( 22\pm 1 \)\\
H\( \gamma \lambda 4340 \)&
\( 37\pm 5 \)&
\( 43\pm 3 \)&
\( 45\pm 1 \)&
\( 51\pm 5 \)&
\( 43\pm 5 \)&
\( 45\pm 2 \)&
\( 47\pm 4 \)\\
{[}OIII{]}\( \lambda 4363 \)&
---&
\( 10\pm 3 \)&
\( 8\pm 1 \)&
\( 5.4\pm 3.2 \)&
\( 10\pm 3 \)&
\( 9\pm 1 \)&
\( 15\pm 4 \)\\
HeII\emph{\( \lambda 4686 \)}&
---&
\( 50\pm 3 \)&
\( 53\pm 1 \)&
\( 47\pm 4 \)&
\( 57\pm 4 \)&
\( 48\pm 1 \)&
\( 50\pm 1 \)\\
({[}ArIV{]})\( \lambda 4713 \)&
---&
\( 6\pm 2 \)&
\( 8\pm 2 \)&
---&
---&
\( 9\pm 2 \)&
\( 8\pm 2 \)\\
H\( \beta \lambda 4861 \)&
\( 100 \)&
\( 100 \)&
\( 100 \)&
\( 100 \)&
\( 100 \)&
\( 100 \)&
\( 100 \)\\
{[}OIII{]}\( \lambda 4959 \)&
\( 283\pm 14 \)&
\( 305\pm 7 \)&
\( 281\pm 4 \)&
\( 301\pm 10 \)&
\( 355\pm 13 \)&
\( 341\pm 6 \)&
\( 352\pm 6 \)\\
{[}OIII{]}\( \lambda 5007 \)&
\( 839\pm 40 \)&
\( 910\pm 21 \)&
\( 867\pm 11 \)&
\( 899\pm 30 \)&
\( 1042\pm 38 \)&
\( 996\pm 16 \)&
\( 1031\pm 18 \)\\
{[}NI{]}\( \lambda 5199 \)&
---&
---&
\( 5\pm 1 \)&
\( <6 \)&
---&
\( 5\pm 1 \)&
\( 4\pm 1 \)\\
{[}FeVII{]}\( \lambda 6087 \)&
---&
---&
---&
---&
---&
\( 14\pm 1 \)&
\( 9\pm 1 \)\\
{[}OI{]}\( \lambda 6300 \)&
---&
\( 10\pm 2 \)&
\( 13\pm 2 \)&
---&
\( 12\pm 3 \)&
\( 11\pm 1 \)&
\( 12\pm 1 \)\\
{[}OI{]}\( \lambda 6364 \)&
---&
\( <4 \)&
\( 4.1\pm 0.6 \)&
---&
---&
\( 4\pm 1 \)&
\( 4\pm 1 \)\\
{[}FeX{]}\( \lambda 6375 \)&
---&
---&
---&
---&
\( 5\pm 2 \)&
\( 4\pm 1 \)&
\( 2\pm 1 \)\\
{[}NII{]}\( \lambda 6548 \)&
\( 43\pm 3 \)&
\( 47\pm 1 \)&
\( 50\pm 1 \)&
\( 52\pm 2 \)&
\( 51\pm 2 \)&
\( 47\pm 1 \)&
\( 36\pm 2 \)\\
H\( \alpha \lambda 6563 \)N&
\( 323\pm 17 \)&
\( 321\pm 8 \)&
\( 270\pm 4 \)&
\( 283\pm 10 \)&
\( 261\pm 10 \)&
\( 278\pm 5 \)&
\( 275\pm 7 \)\\
H\( \alpha \lambda 6563 \)B&
---&
---&
\( 25\pm 10 \)&
---&
\( 58\pm 10 \)&
\( 66\pm 9 \)&
\( 57\pm 15 \)\\
{[}NII{]}\( \lambda 6583 \)&
\( 130\pm 8 \)&
\( 140\pm 4 \)&
\( 150\pm 3 \)&
\( 155\pm 7 \)&
\( 151\pm 6 \)&
\( 141\pm 3 \)&
\( 108\pm 5 \)\\
HeI\( \lambda 6679 \)&
---&
\( <2 \)&
\( 6\pm 2 \)&
\( <2 \)&
---&
\( 6\pm 2 \)&
\( 4\pm 1 \)\\
{[}SII{]}\( \lambda 6716 \)&
\( 69\pm 10 \)&
\( 46\pm 2 \)&
\( 42\pm 1 \)&
\( 40\pm 3 \)&
\( 40\pm 2 \)&
\( 42\pm 1 \)&
\( 36\pm 1 \)\\
{[}SII{]}\( \lambda 6731 \)&
\( 54\pm 7 \)&
\( 31\pm 2 \)&
\( 40\pm 1 \)&
\( 40\pm 3 \)&
\( 36\pm 2 \)&
\( 35\pm 1 \)&
\( 31\pm 1 \)\\
{[}ArIII{]}\( \lambda 7136 \)&
---&
\( 16\pm 2 \)&
\( 16\pm 1 \)&
\( 19\pm 3 \)&
\( 23\pm 4 \)&
\( 17\pm 1 \)&
\( 18\pm 1 \)\\
\hline 
\end{tabular}\par}

\caption{\label{tab: Reddening corrected EL} As Table \ref{tab: EL fluxes} above,
but corrected for reddening using the values of E(B-V) calculated for each region.
Note the consistency across the different spatial regions for any given line
ratio.}

\( ^{a} \)Values for blends.

\( ^{b} \)Includes contribution from H\( \epsilon  \)
\end{table}

\begin{table}[p]
{\centering \begin{tabular}{|c|c|c|c|c|c||c|}
\hline 
&
\( T \)&
\( d^{\dagger } \)&
\( \mbox {n}_{e} \)&
\( U \)&
\( Q \)&
\( Q \) (\( U=5\times 10^{-4} \))\\
&
(K)&
(\( 10^{20} \)m)&
(\( \mbox {cm}^{-3} \))&
(\( 10^{-2} \))&
(phot \( \mbox {s}^{-1}\mbox {sr}^{-1} \))&
\\
\hline 
A&
---&
\( 5.4\pm 0.7 \)&
\( <380 \)&
\( 0.56\pm 0.04 \)&
---&
---\\
B&
\( 12000_{-1000}^{+1900} \)&
\( 4.0\pm 0.7 \)&
\( <60 \)&
\( 1.20\pm 0.05 \)&
---&
---\\
C&
\( 11000_{-300}^{+600} \)&
\( 2.5\pm 0.6 \)&
\( 500_{-70}^{+60} \)&
\( 0.95\pm 0.02 \)&
\( (8.8\pm 2.4)\times 10^{55} \)&
\( (4.7\pm 1.2)\times 10^{54} \)\\
D&
\( 10000_{-1300}^{+4000} \)&
\( 1.4\pm 0.6 \)&
\( 550_{-180}^{+240} \)&
\( 0.71\pm 0.04 \)&
\( (4.2\pm 2.4)\times 10^{55} \)&
\( (1.5\pm 0.9)\times 10^{54} \)\\
E&
\( 11600_{-900}^{+1700} \)&
\( 0.4\pm 0.3 \)&
\( 390_{-130}^{+170} \)&
\( 0.98_{-0.05}^{+0.06} \)&
\( (2.1\pm 1.9)\times 10^{54} \)&
\( (1.1\pm 1.0)\times 10^{53} \)\\
F&
\( 11300\pm 600 \)&
\( 0.4\pm 0.5 \)&
\( 260\pm 40 \)&
\( 1.05\pm 0.03 \)&
\( (1.1\pm 1.4)\times 10^{54} \)&
\( (5.3\pm 6.6)\times 10^{52} \)\\
G&
\( 13300_{-1100}^{+1600} \)&
\( 1.3\pm 0.5 \)&
\( 310\pm 60 \)&
\( 1.3_{-0.03}^{+0.04} \)&
\( (2.1\pm 0.8)\times 10^{55} \)&
\( (7.7\pm 3.1)\times 10^{53} \)\\
\hline 
\end{tabular}\par}

\caption{\label{tab: Physical conditions}Table showing reddening corrected values of
temperature, distance from `nucleus,' electron density, ionization parameter
and photoionizing luminosity, respectively. The upper limits for the density
in regions A \& B are \protect\( 1\sigma \protect \) limits. The uncertainties
in \protect\( Q\protect \) are dominated by the uncertainties in the distances,
which correspond to the width of the regions A-G. Also shown is the value of
\protect\( Q\protect \) derived for each region assuming a value of \protect\( U\protect \)
more appropriate to the mixed medium models discussed in Section \ref{sec: Photoionization models}.}

\( ^{\dagger } \)Assumes H\( _{\circ }=50\: \mbox {kms}^{-1}\mbox {Mpc}^{-1} \),
\( q_{\circ }=0.5 \)
\end{table}

\newpage
\begin{figure}[p]

\caption{(See accompanying jpeg file.) Position of the slit superimposed on an {[}OIII{]}\protect\( \lambda \lambda 4959+5007\protect \)
image taken using the HST. The different regions for which the various measurements
and calculations detailed in Section 3 were carried out are labelled A-G. Note
that there is a gap between regions B \& C \protect\cite{Martel+B+S+etal-1999}.\label{fig: HST Image}}
\end{figure}

\begin{figure}[p]
{\par\centering \resizebox*{1\textwidth}{!}{\includegraphics{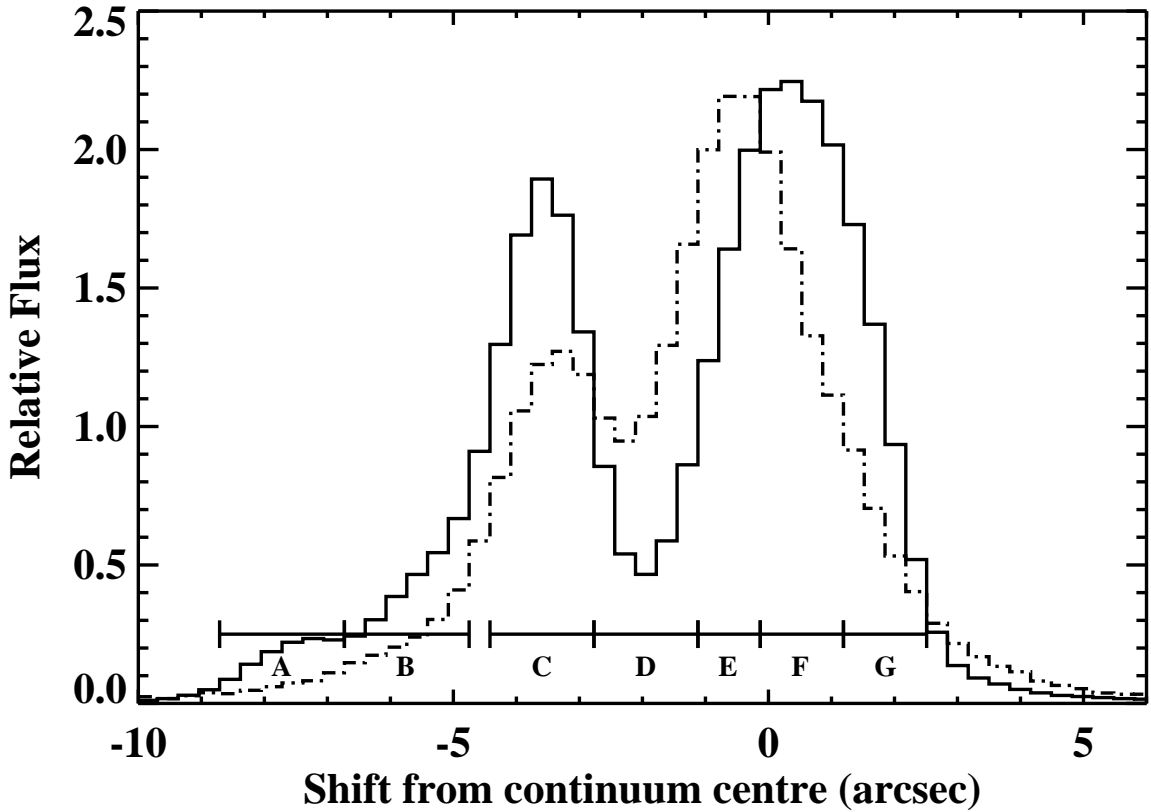}} \par}

\caption{\label{fig: Bins}Plot showing the spatial variation of {[}OIII{]}\protect\( \lambda 5007\protect \)
emission (solid line), together with that of a continuum bin centred on 5110
\AA ~(dashed line) in the rest frame. The positions of the spatial bins chosen
for analysis are also shown. The continuum flux has been scaled by a factor
of \protect\( \sim 50\protect \) for ease of comparison. `0' is defined as
the position of the continuum measured by fitting a Gaussian to the continuum
spatial profile in a bin centred on 8849 \AA.}
\end{figure}

\begin{figure}[p]
{\par\centering \resizebox*{13cm}{!}{\includegraphics{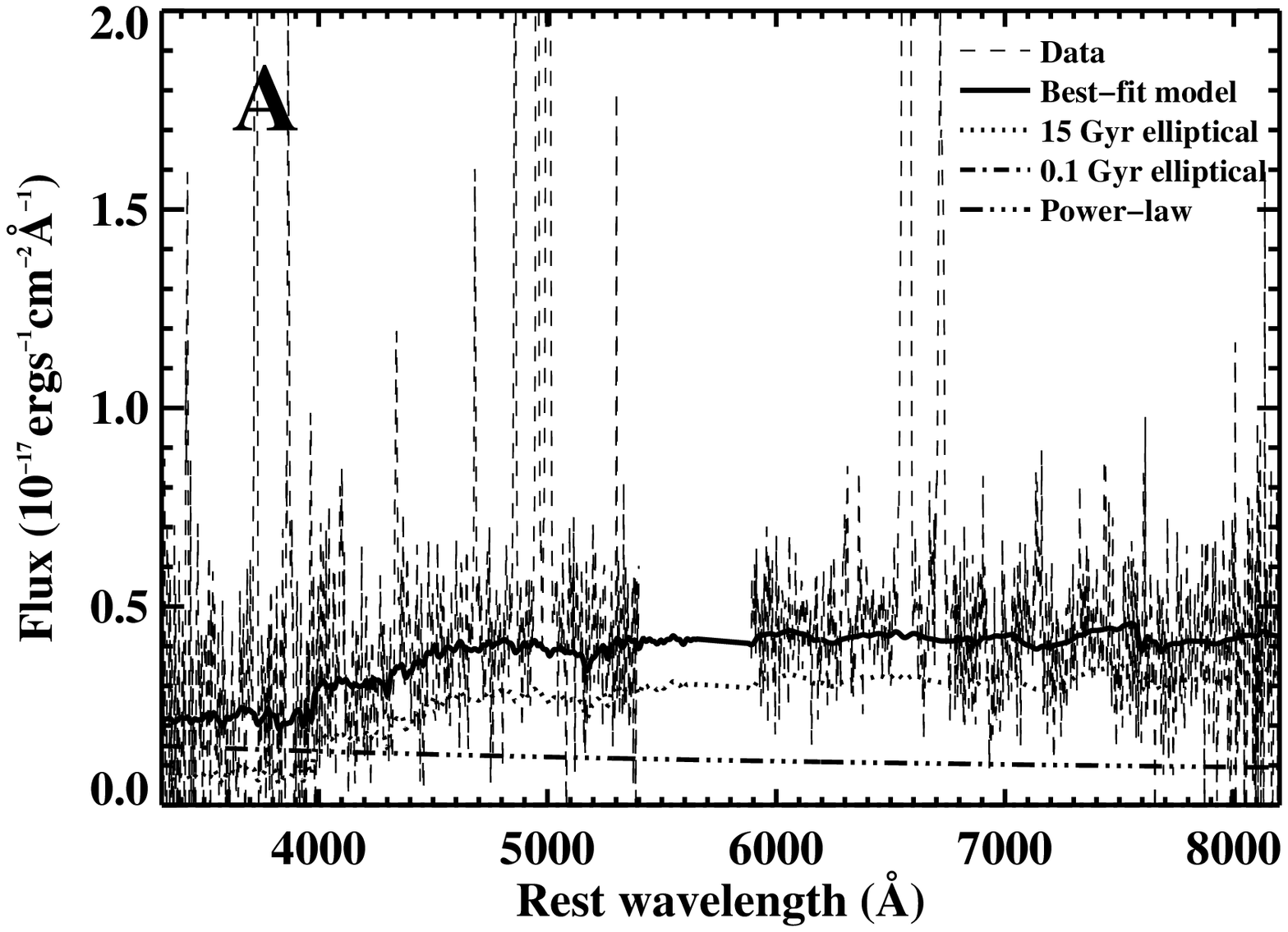}} \par}

{\par\centering \resizebox*{13cm}{!}{\includegraphics{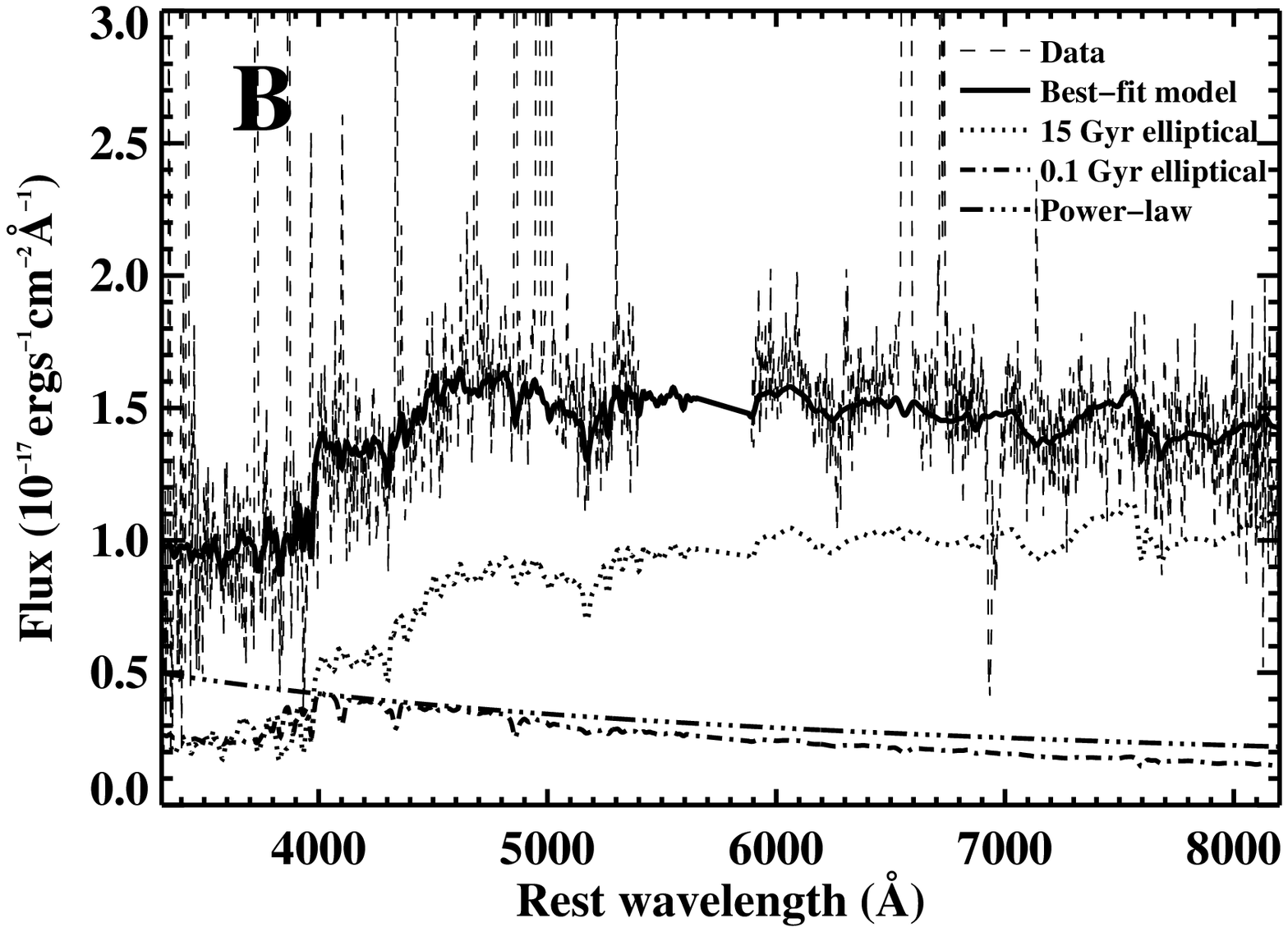}} \par}

\caption{\label{fig: Continuum subtracted spectra}ISIS spectra of the different regions
of 3C 321, together with best-fit composite 3-component model spectra and individual
model components. Individual plot legends provide the key to the components
displayed. The 3-component model which minimizes \protect\( \chi ^{2}\protect \)
in each case can be seen to fit the continuum in the different regions extremely
well.}
\end{figure}

\begin{figure*}[p]
{\par\centering \resizebox*{13cm}{!}{\includegraphics{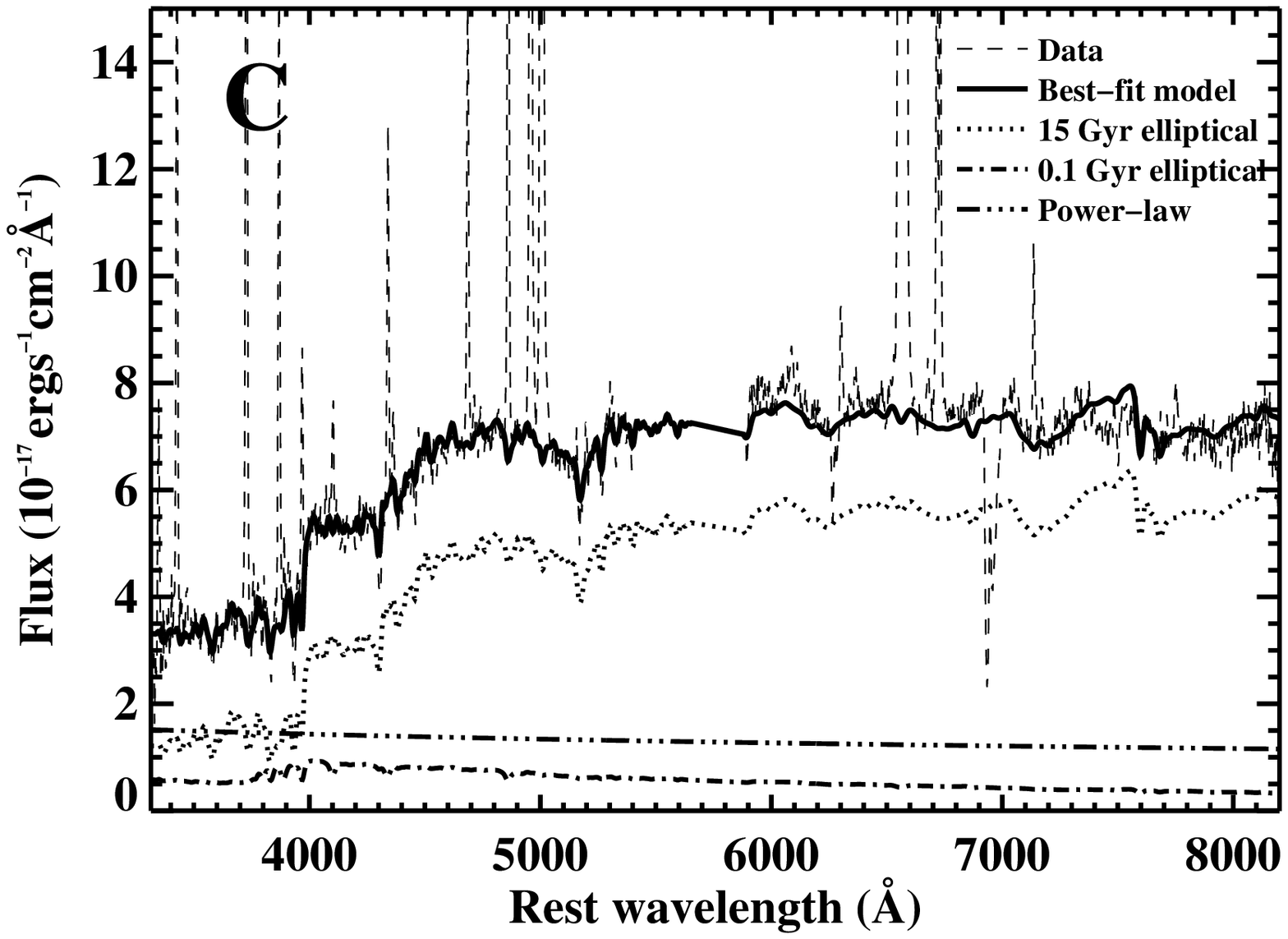}} \par}

{\par\centering \resizebox*{13cm}{!}{\includegraphics{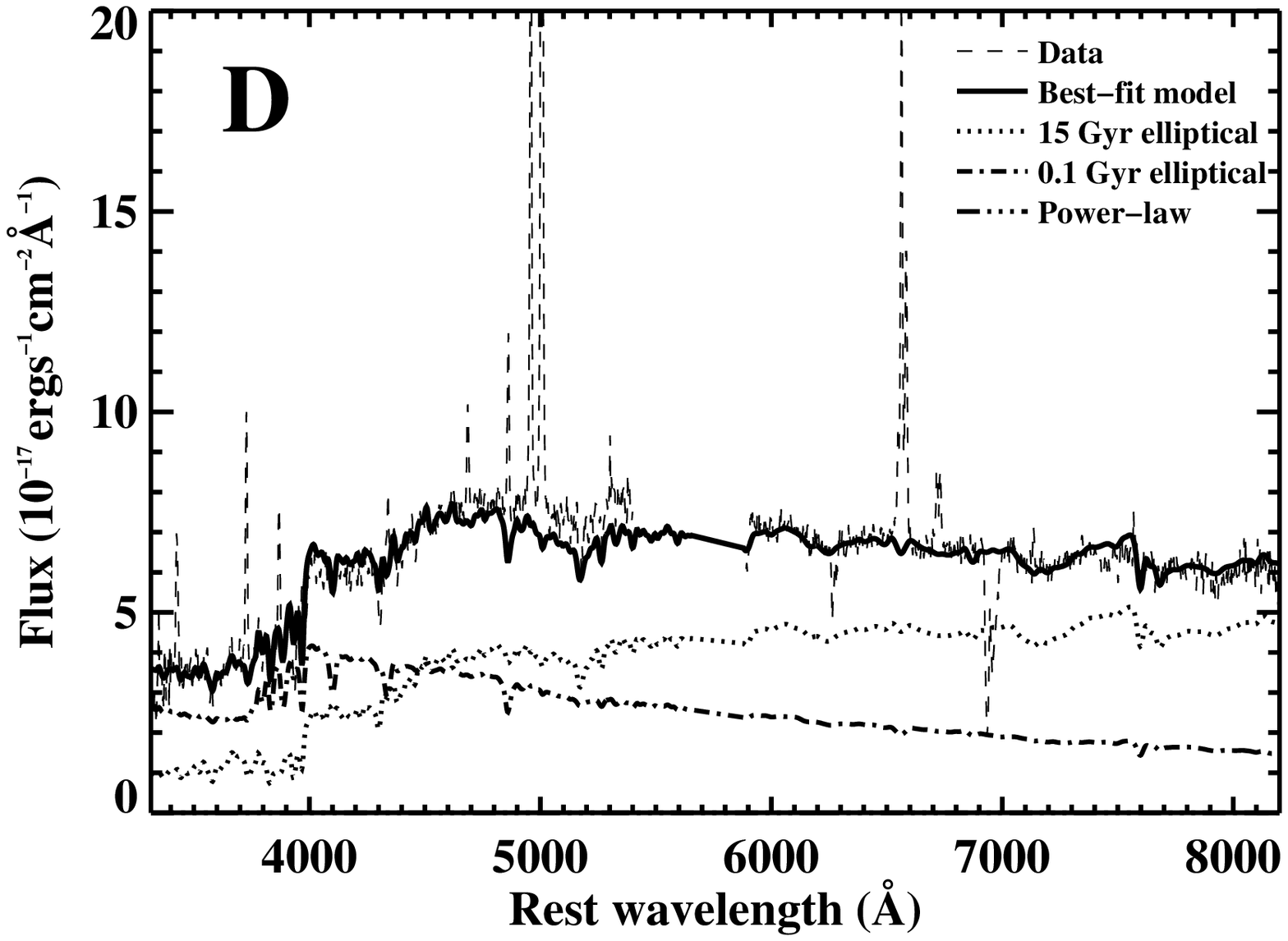}} \par}

\dittocaption{Continued}
\end{figure*}

\begin{figure*}[p]
{\par\centering \resizebox*{13cm}{!}{\includegraphics{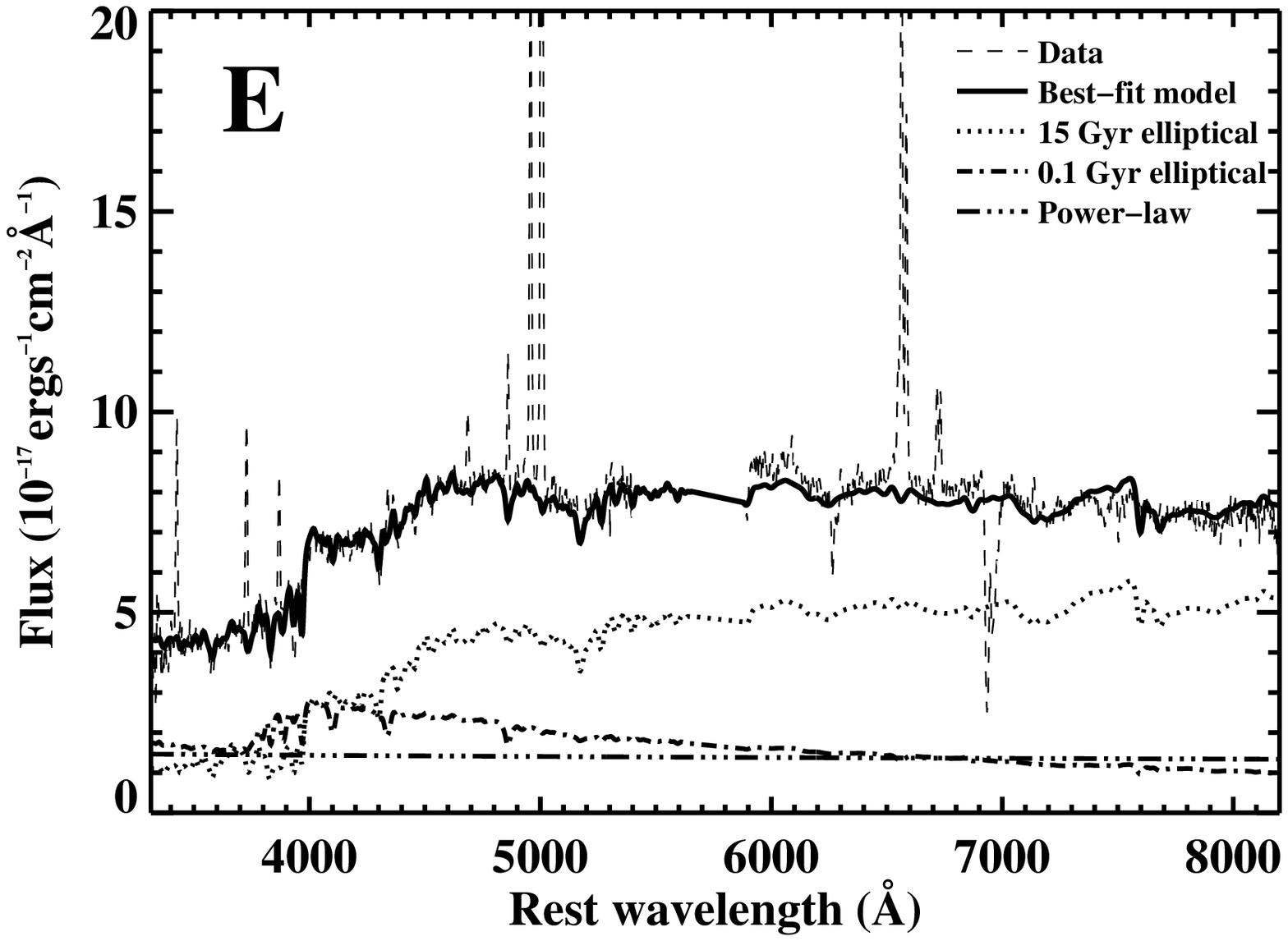}} \par}

{\par\centering \resizebox*{13cm}{!}{\includegraphics{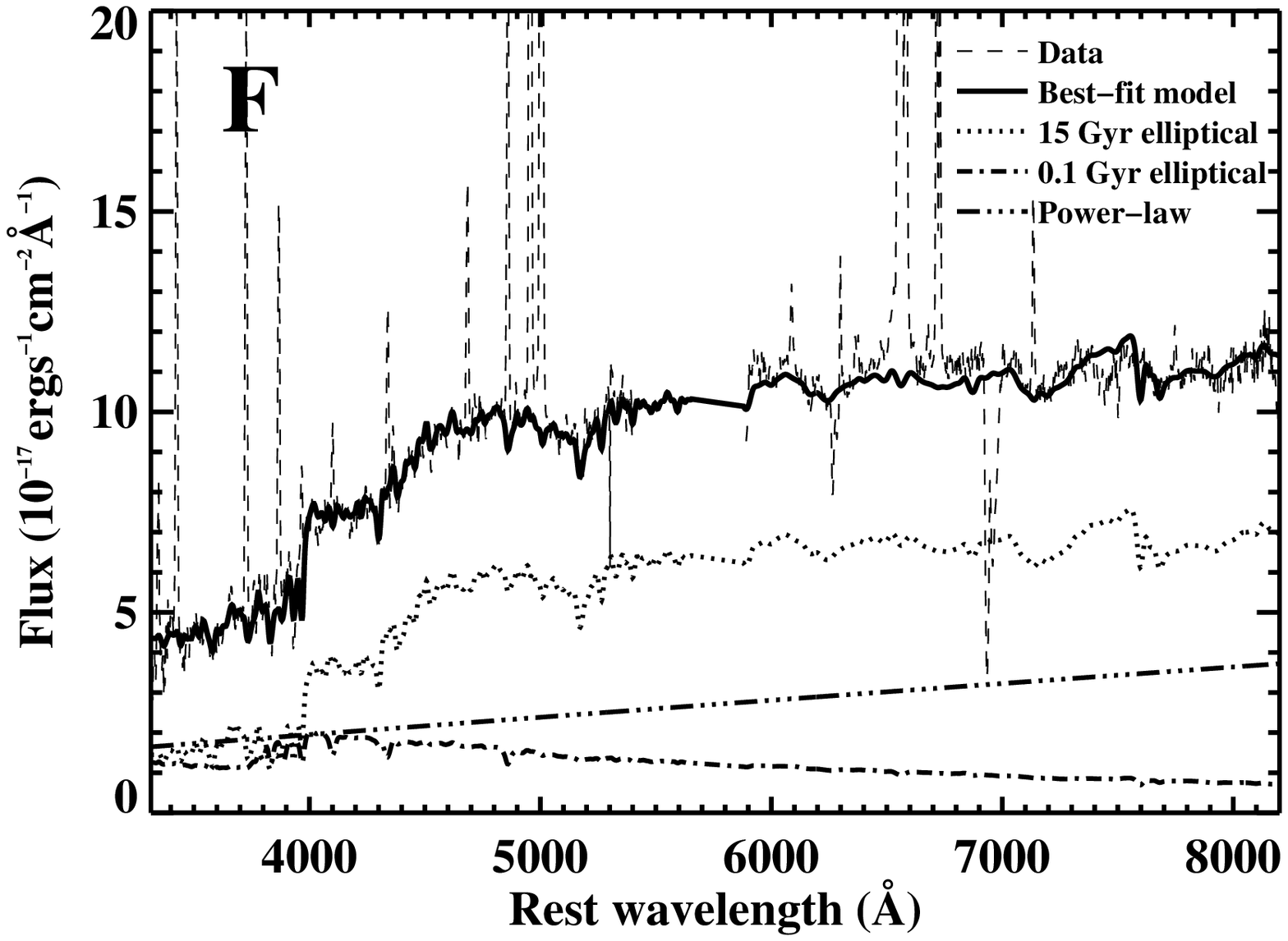}} \par}

\dittocaption{Continued}
\end{figure*}

\begin{figure*}[p]
{\par\centering \resizebox*{13cm}{!}{\includegraphics{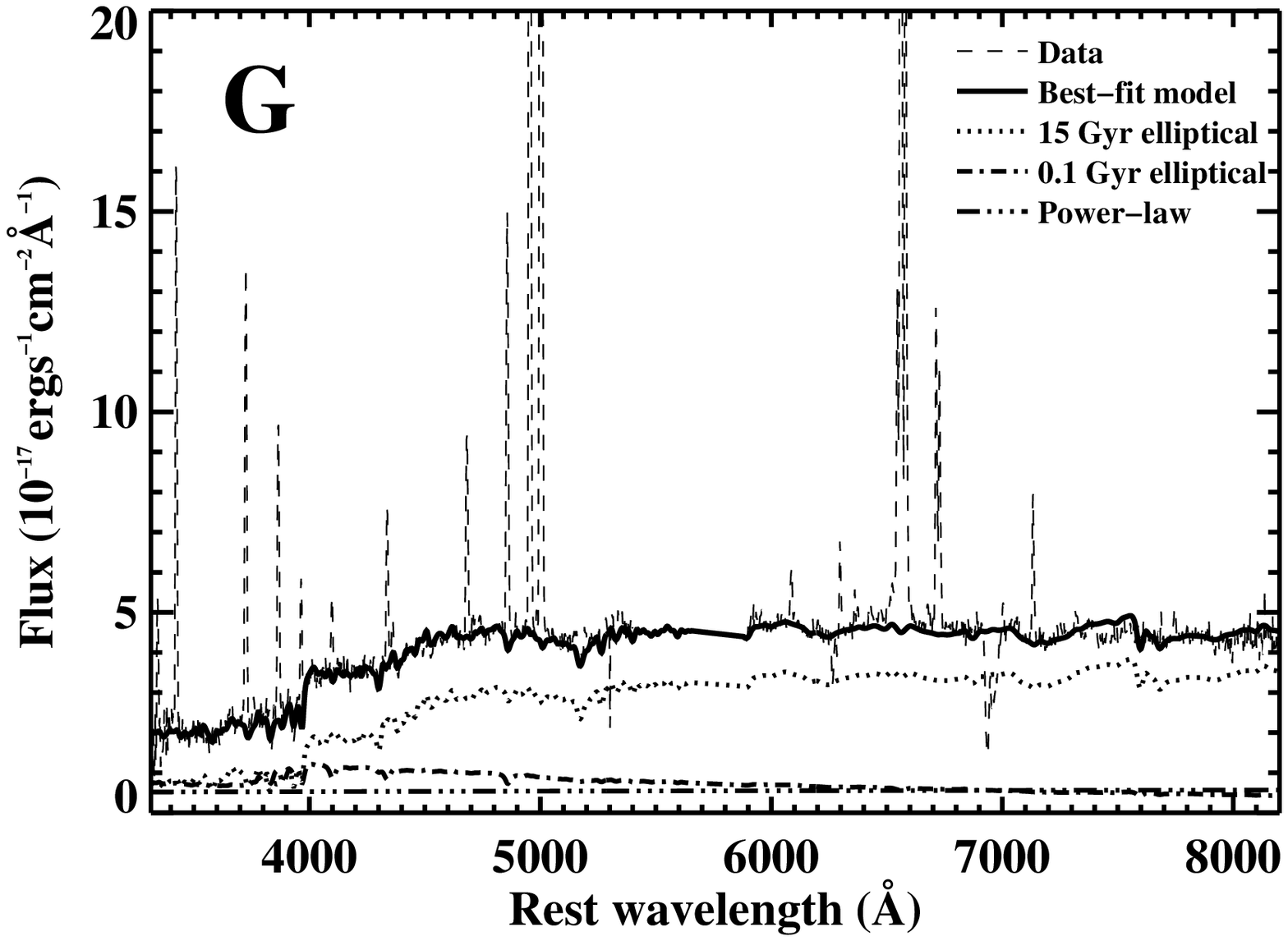}} \par}

\dittocaption{Continued}
\end{figure*}

\begin{figure}[p]
{\par\centering \subfigure[ ]{\resizebox*{!}{0.4\textheight}{\includegraphics{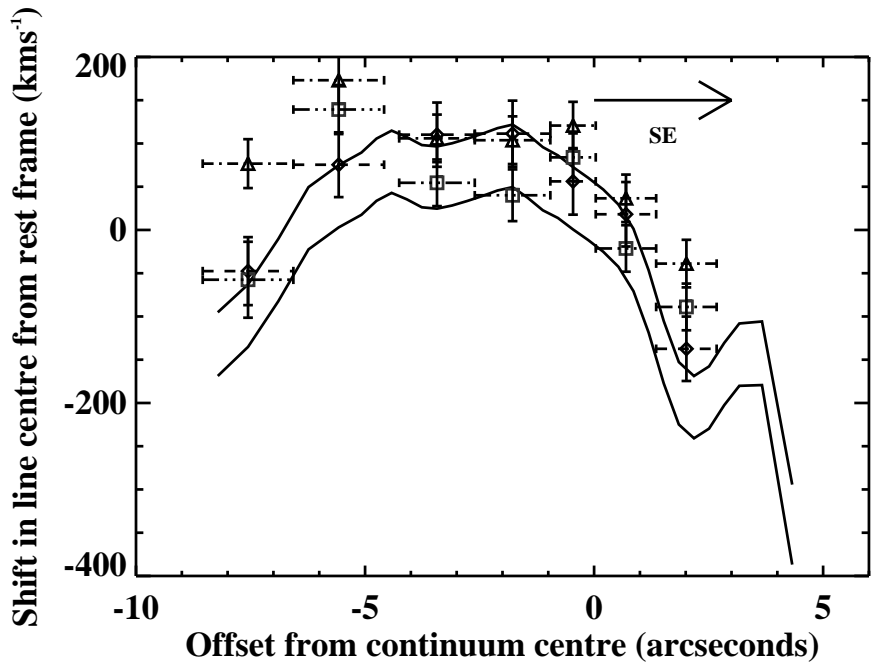}}} \par}

{\par\centering \subfigure[ ]{\resizebox*{!}{0.4\textheight}{\includegraphics{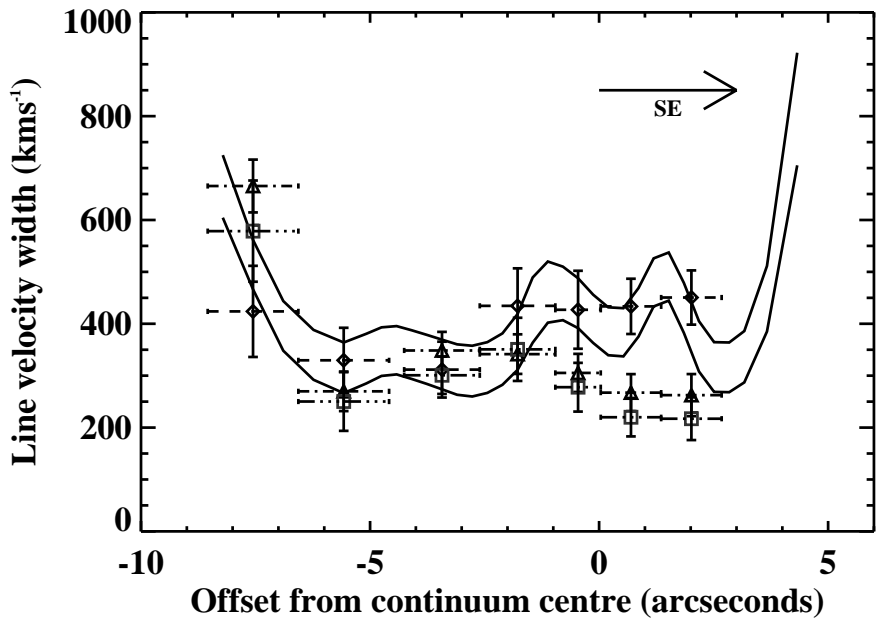}}} \par}

\caption{\label{fig: EL kinematics} Emission-line kinematics. Plots tracing the variation
in a) line centre and b) line width across 3C 321. The solid lines denote the
1 sigma error bars on {[}OIII{]}\protect\( \lambda 5007\protect \) measured
on a pixel-by-pixel basis across the image. The triangles denote H\protect\( \alpha \protect \),
diamonds H\protect\( \beta \protect \) and squares {[}SII{]}\protect\( \lambda 6717\protect \). }
\end{figure}

\begin{figure}[p]
a)

{\par\centering \resizebox*{13cm}{!}{\includegraphics{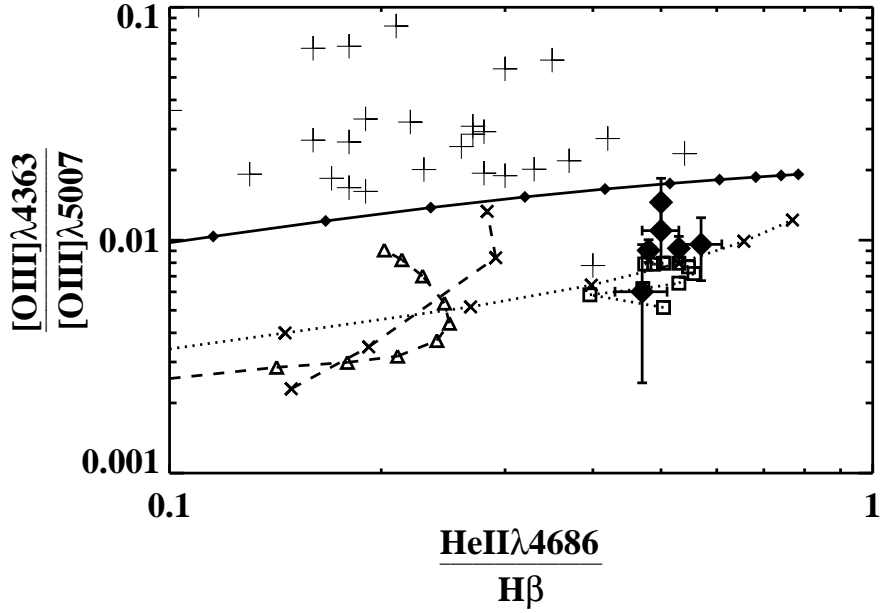}} \par}

b)

{\par\centering \resizebox*{13cm}{!}{\includegraphics{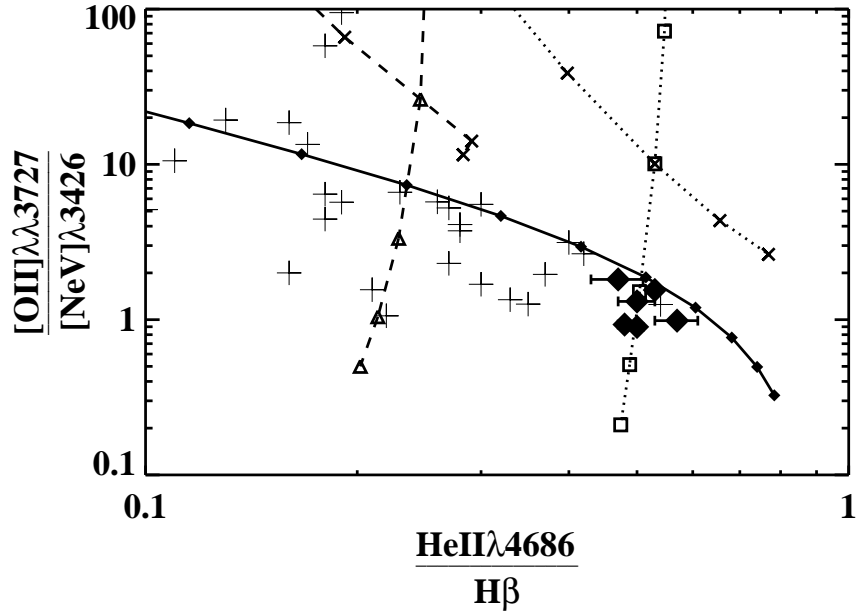}}  \par}

\caption{\label{fig: Diagdiags}Diagnostic diagrams showing a) {[}OIII{]}\protect\( \lambda 4363/5007\protect \)
vs HeII\protect\( \lambda 4686\protect \)/H\protect\( \beta \protect \), b)
{[}OII{]}\protect\( \lambda \lambda 3727\protect \)/{[}NeV{]}\protect\( \lambda 3426\protect \)
vs HeII\protect\( \lambda 4686\protect \)/H\protect\( \beta \protect \), c)
{[}NII{]}\protect\( \lambda 6583\protect \)/{[}OII{]}\protect\( \lambda \lambda 3727\protect \)
vs {[}OII{]}\protect\( \lambda \lambda 3727\protect \)/{[}OIII{]}\protect\( \lambda 5007\protect \),
d) {[}OIII{]}\protect\( \lambda 4363/5007\protect \) vs {[}OII{]}\protect\( \lambda \lambda 3727\protect \)/{[}OIII{]}\protect\( \lambda 5007\protect \),
e) {[}NeIII{]}\protect\( \lambda 3869\protect \)/{[}NeV{]}\protect\( \lambda 3426\protect \)
vs HeII/H\protect\( \beta \protect \), f) {[}OII{]}\protect\( \lambda \lambda 3727\protect \)/{[}OIII{]}\protect\( \lambda 5007\protect \)
vs {[}OI{]}\protect\( \lambda 6300\protect \)/{[}OIII{]}\protect\( \lambda 5007\protect \),
g) {[}OIII{]}\protect\( \lambda 5007\protect \)/H\protect\( \beta \protect \)
vs {[}OI{]}\protect\( \lambda 6300\protect \)/{[}OIII{]}\protect\( \lambda 5007\protect \),
h) {[}SII{]}\protect\( \lambda 6724\protect \)/H\protect\( \alpha \protect \)
vs {[}OII{]}\protect\( \lambda \lambda 3727\protect \)/{[}OIII{]}\protect\( \lambda 5007\protect \),
i) {[}OIII{]}\protect\( \lambda 5007\protect \)/H\protect\( \beta \protect \)
vs {[}OII{]}\protect\( \lambda \lambda 3727\protect \)/{[}OIII{]}\protect\( \lambda 5007\protect \).
Dashed lines represent power-law ionizing continua, \protect\( f\propto \nu ^{\alpha _{\nu }}\protect \),
with triangles representing \protect\( \alpha _{\nu }=-1.5,\protect \) \protect\( U=10^{-4},\protect \)
\protect\( 2.5\times 10^{-4},\protect \) \protect\( 5\times 10^{-4},\protect \)
\protect\( 10^{-3},\protect \) \protect\( 2.5\times 10^{-3},\protect \) \protect\( 5\times 10^{-3},\protect \)
\protect\( 0.01,\protect \) \protect\( 0.025,\protect \) \protect\( 0.05,\protect \)
\protect\( 0.1\protect \) and crosses (\protect\( \times \protect \)) showing
models with \protect\( U\protect \) fixed at \protect\( 0.01\protect \) and
\protect\( \alpha _{\nu }=-2.0,\protect \) \protect\( -1.75,\protect \) \protect\( -1.5,\protect \)
\protect\( -1.0\protect \). Dotted lines are black body models, with squares
representing sequences of models with \protect\( T=160,000\protect \) K and
the same sequence in \protect\( U\protect \) as before, and crosses showing
sequences with \protect\( U=0.01\protect \) and \protect\( T=(0.6,\protect \)
\protect\( 0.8,\protect \) \protect\( 1.0,\protect \) \protect\( 1.2,\protect \)
\protect\( 1.4,\protect \) \protect\( 1.6,\protect \) \protect\( 1.8,\protect \)
\protect\( 2.0)\times 10^{5}\protect \) K. Solid lines correspond to models
with a \protect\( \alpha _{\nu }=-1.3\protect \) power law illuminating mixture
of optically thin and thick clouds. The reddening corrected values for 3C 321
are shown as filled diamonds. `+' denotes measurements for other radio galaxies
taken from \protect\shortcite{Dickson-1997}, \protect\shortcite{Tadhunter-1983}, \protect\shortcite{Grandi+O-1978}, \protect\shortcite{Costero+O-1977},
\protect\shortcite{Koski-1978}, \protect\shortcite{Cohen+O-1981} and \protect\shortcite{Tadhunter+M+R-1994}.}
\end{figure}

\begin{figure}[p]
c)

{\par\centering \resizebox*{13cm}{!}{\includegraphics{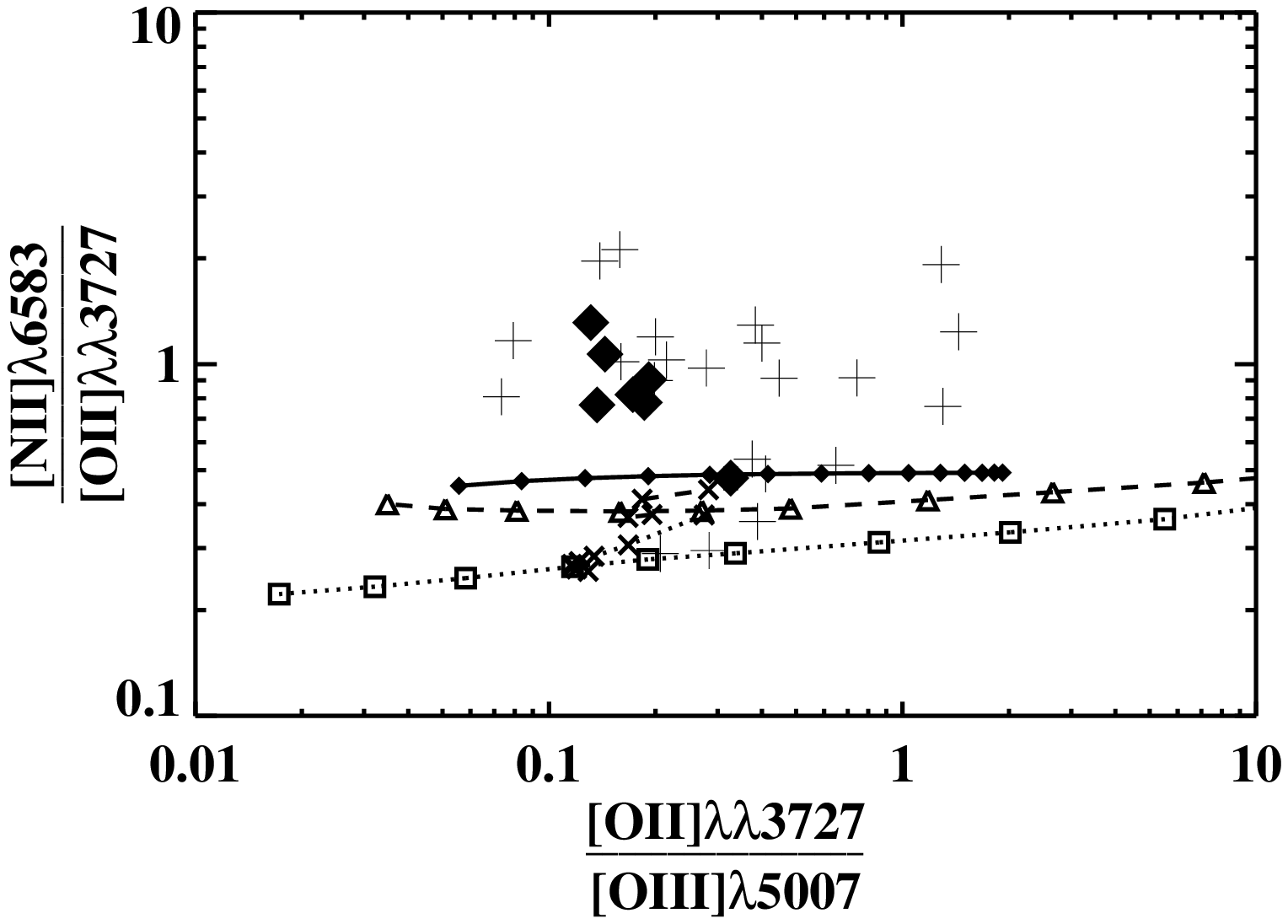}} \par}

d)

{\par\centering \resizebox*{13cm}{!}{\includegraphics{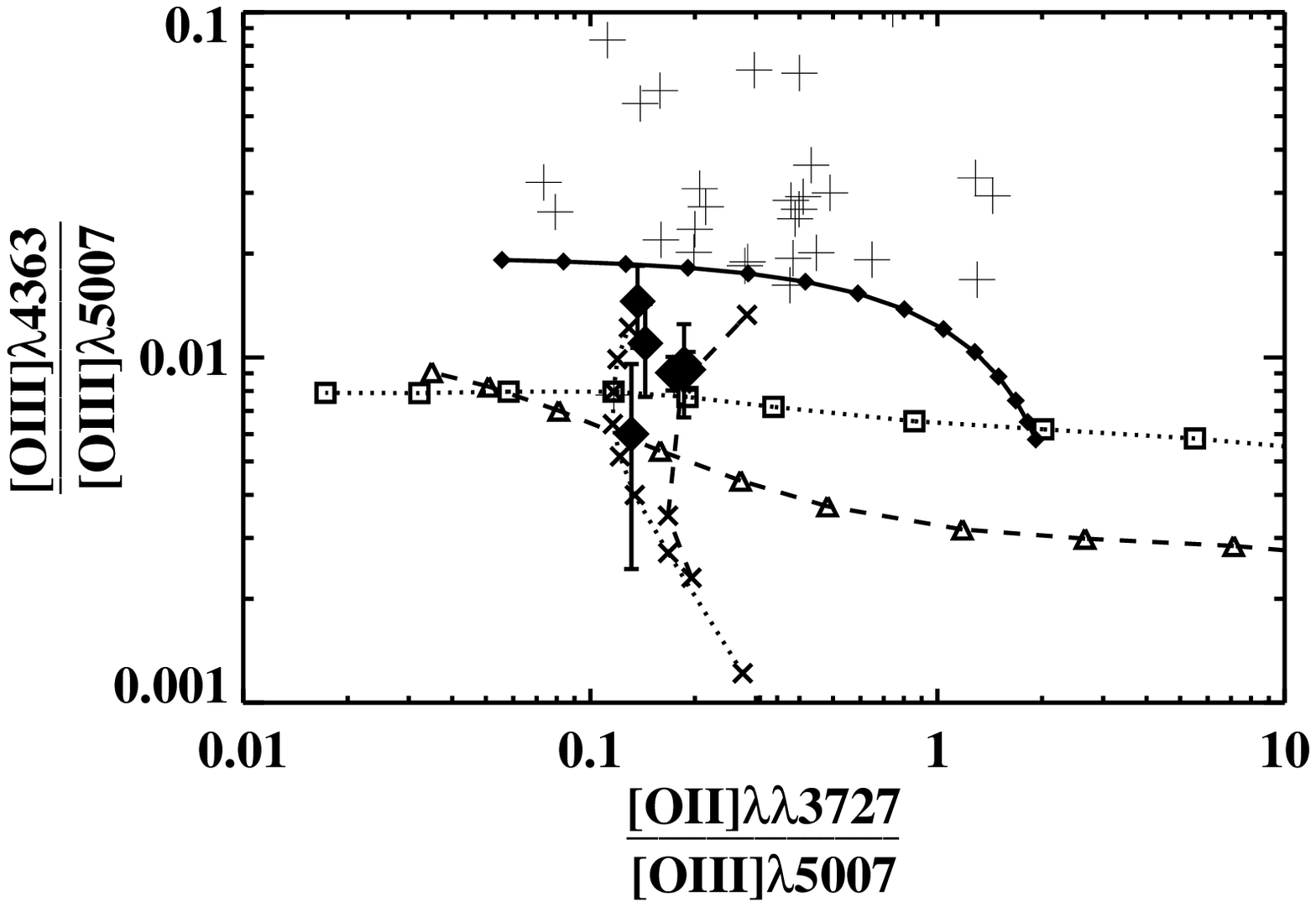}} \par}

\dittocaption{continued}
\end{figure}

\begin{figure}[p]
e)

{\par\centering \resizebox*{13cm}{!}{\includegraphics{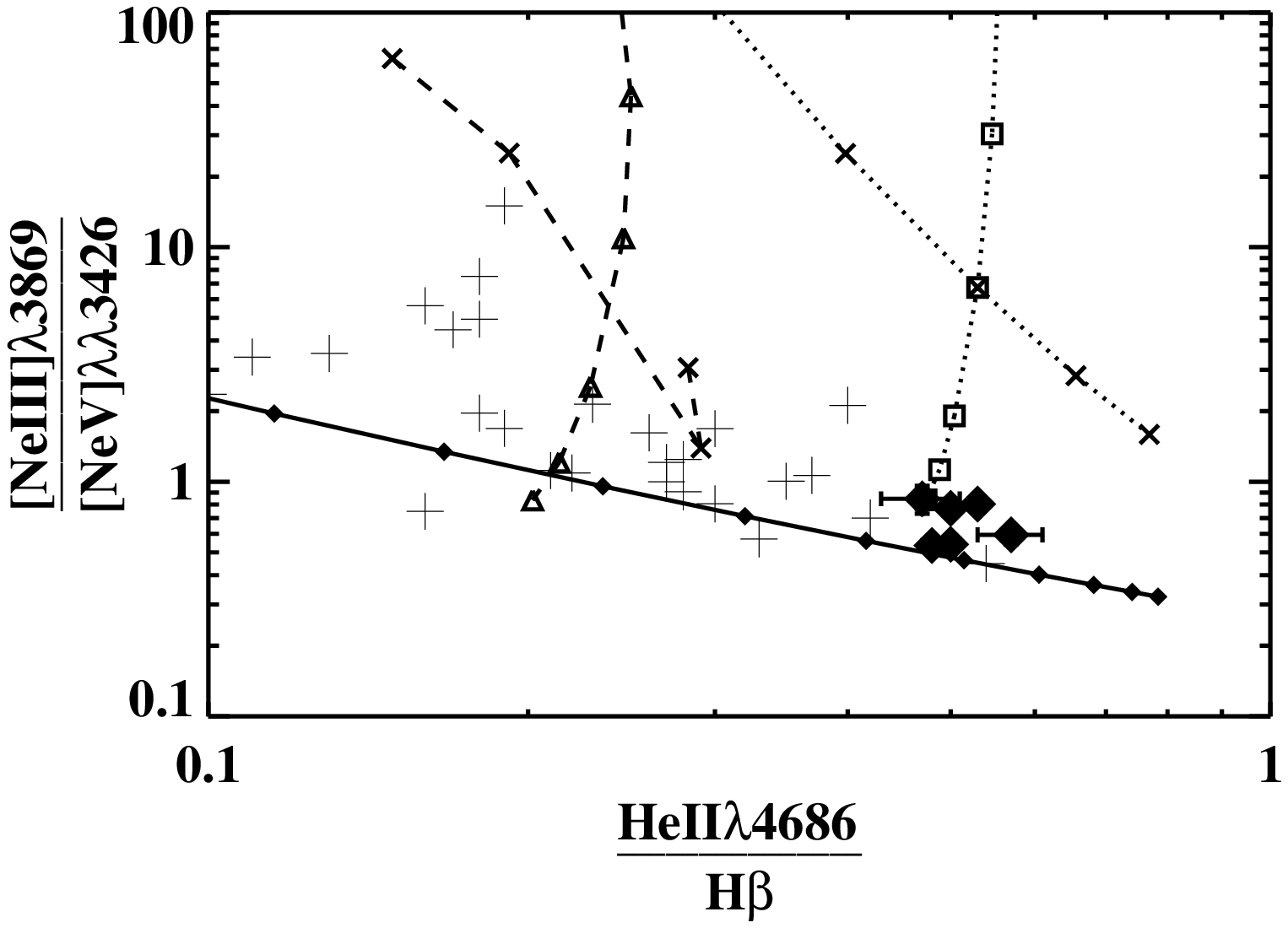}} \par}

f)

{\par\centering \resizebox*{13cm}{!}{\includegraphics{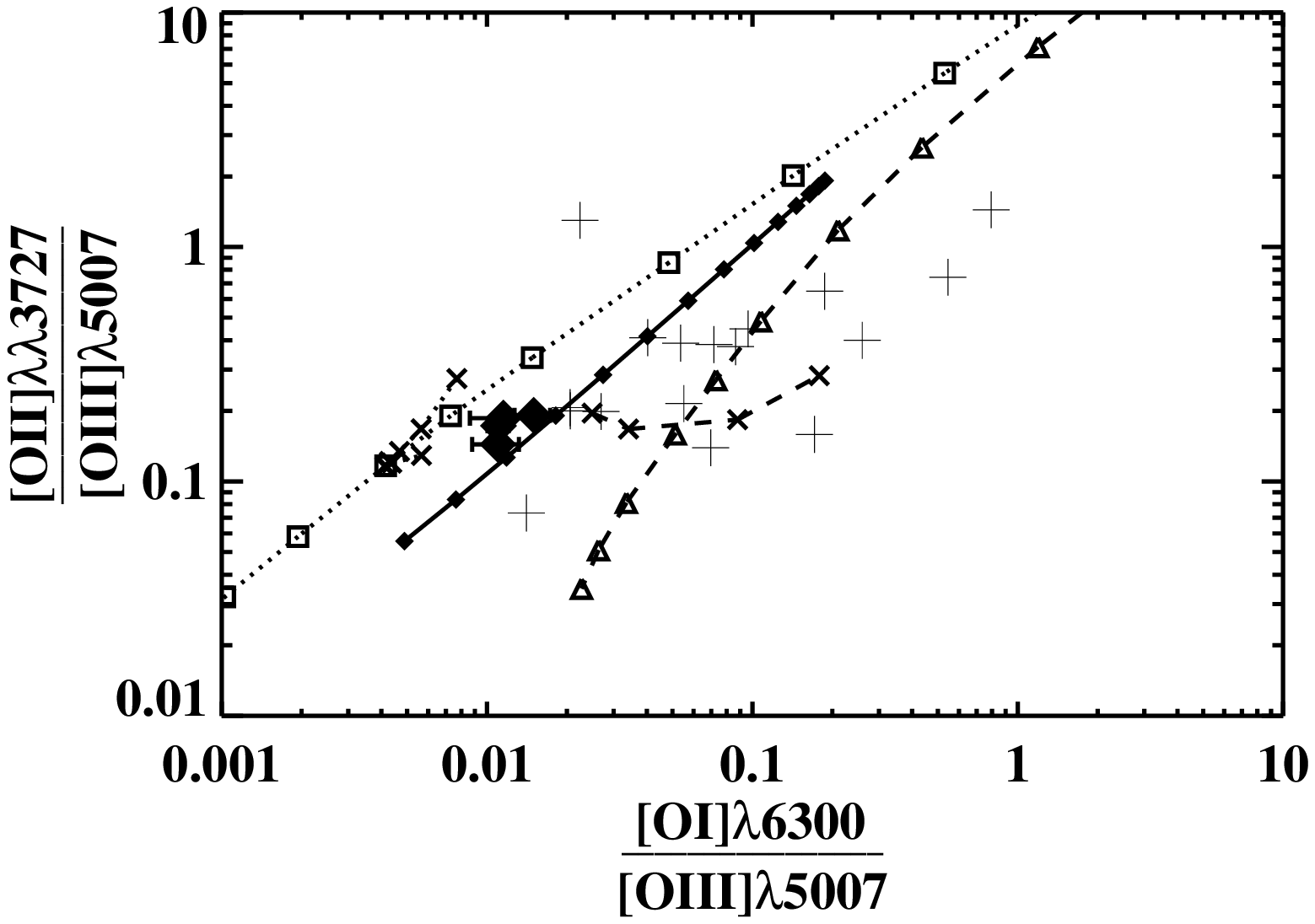}} \par}

\dittocaption{continued}
\end{figure}

\begin{figure}[p]
g)

{\par\centering \resizebox*{13cm}{!}{\includegraphics{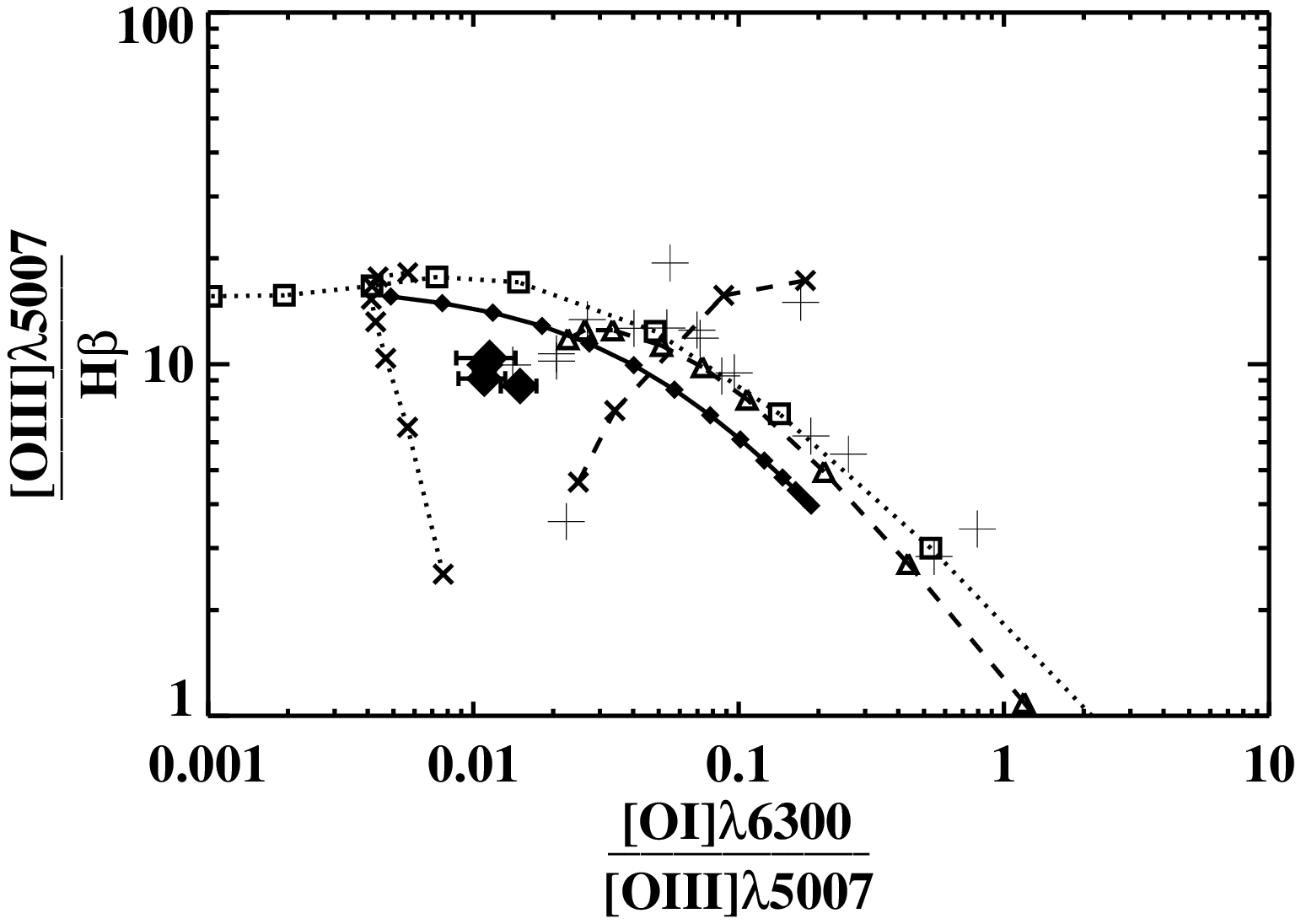}} \par}

h)

{\par\centering \resizebox*{13cm}{!}{\includegraphics{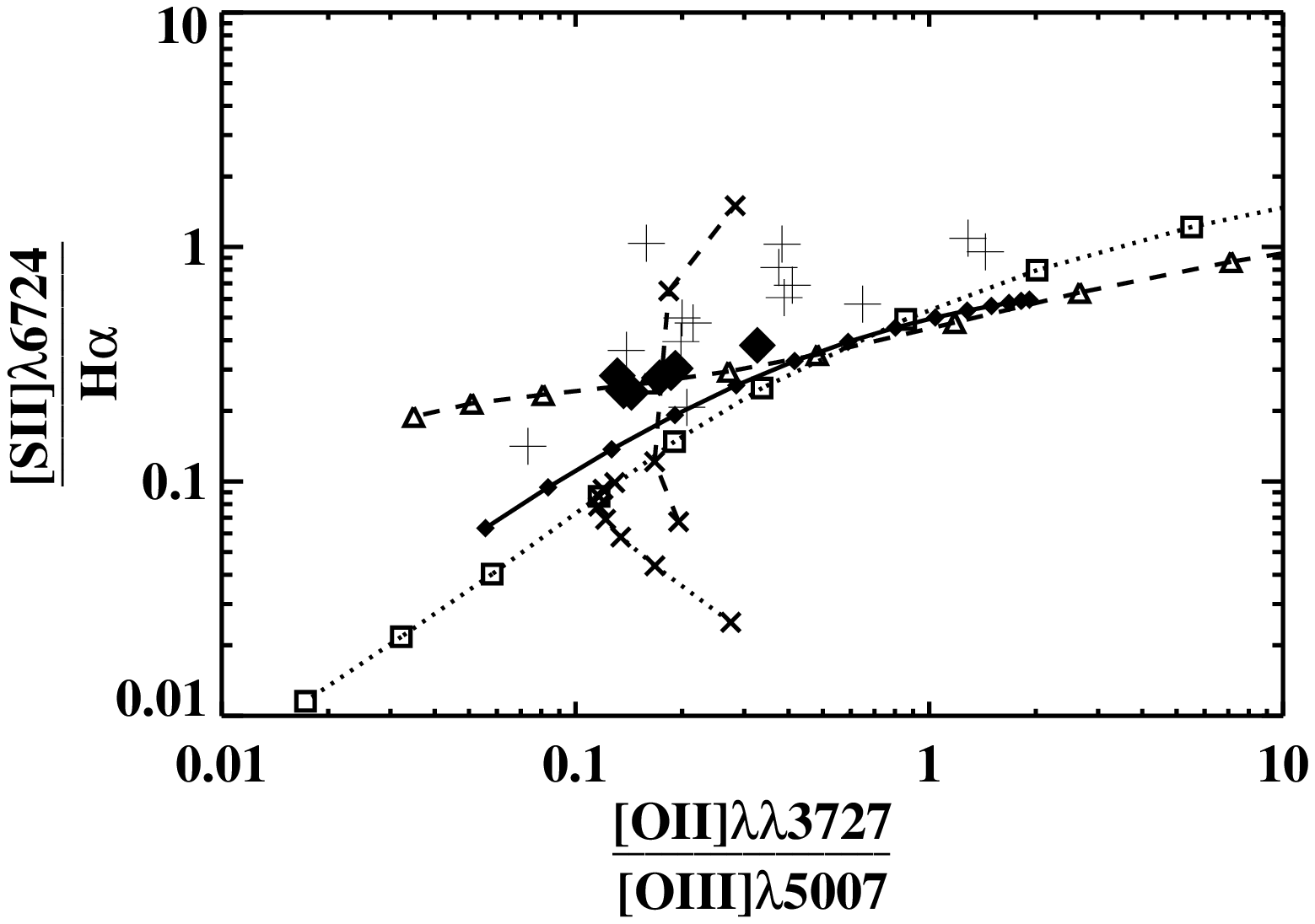}} \par}

\dittocaption{continued}
\end{figure}

\begin{figure}[p]
i)

{\par\centering \resizebox*{13cm}{!}{\includegraphics{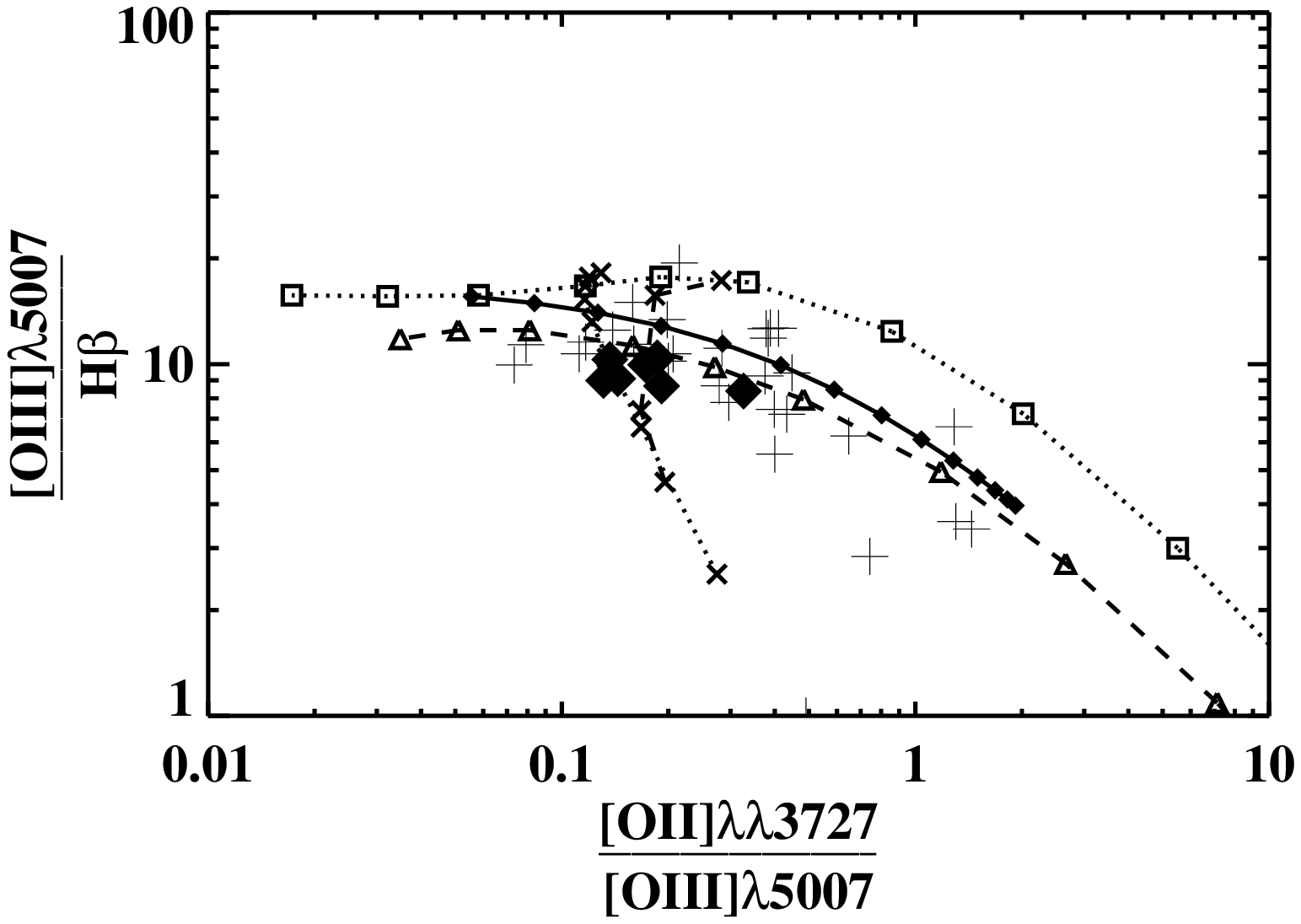}} \par}

\dittocaption{continued}
\end{figure}

\end{document}